\documentclass[useAMS,usenatbib]{mn2e}
\pdfoutput=1
\usepackage{graphicx}
\usepackage{natbib}     
\usepackage{amsmath} 
\usepackage{float} 
\usepackage{amssymb}    
\usepackage{footnote}
\usepackage{longtable}
\title[H$\alpha$ and $u$-band accretion rates in the Lagoon Nebula]{Classical T\,Tauri stars with VPHAS$+$: I: H$\alpha$ and $u$-band accretion rates in the Lagoon Nebula M8}
\author[V.~M.~Kalari~et~al.]{V.~M.~Kalari$^{1,\,2\,}$\thanks{E-mail:
:\,vek@arm.ac.uk}, J.~S.~Vink$^{1}$, J.~E.~Drew$^{3}$, G.~Barentsen$^{3}$, J.~J.~Drake$^{4}$, J.~Eisl\"{o}ffel$^{5}$,\newauthor
E.~L.~{Mart{\'{\i}}n}$^{6}$, Q.~A.~Parker$^{7,8}$, Y.~C.~Unruh$^{9}$, N.~A.~Walton$^{10}$, N.~J.~Wright$^{3}$. \\
$^{1}$Armagh Observatory, College Hill, Armagh, BT61\,9DG, U.K.\\
$^{2}$School of Mathematics \& Physics, Queen's University Belfast, Belfast, BT61\,7NN, U.K.\\
$^{3}$Centre for Astronomy Research, Science and Technology Research Institute, University of Hertfordshire, Hatfield, AL10\,9AB, U.K.\\
$^{4}$Harvard-Smithsonian Center for Astrophysics, 60 Garden Street, Cambridge, MA\,02138, U.S.A.\\
$^{5}$Th\"{u}ringer Landessternwarte, Sternwarte 5, 07778, Tautenburg, Germany\\
$^{6}$Departamento de Astrofisica, Centro de Astrobiologia, Carretera de Ajalvir km 4, 28550 Torrejon de Ardoz, Madrid, Spain\\
$^{7}$Department of Physics, Hong Kong University, Hong Kong\\
$^{8}$Australian Astronomical Observatory, P.O. Box\,915, North Ryde New South Wales 1670, Australia\\
$^{9}$Department of Physics, Blackett Laboratory, Imperial College London, Prince Consort Road, London, SW7\,2AZ, U.K.\\
$^{10}$Institute of Astronomy, Cambridge University, Madingley Road, Cambridge, CB3\,OHA, U.K.}
\begin{document}
    
\date{16 December 2014}

\pagerange{\pageref{firstpage}--\pageref{lastpage}} \pubyear{2014}

\maketitle

\label{firstpage}

\begin{abstract}
We estimate the accretion rates of 235 Classical T\,Tauri star (CTTS) candidates in the Lagoon Nebula using {\it{ugri}}H$\alpha$ photometry from the VPHAS+ survey. Our sample consists of stars displaying H$\alpha$-excess, the intensity of which is used to derive accretion rates.
For a subset of 87 stars, the intensity of the $u$-band excess is also used to estimate accretion rates. We find the mean variation in accretion rates measured using H$\alpha$ and $u$-band intensities to be $\sim$\,0.17\,dex, agreeing with previous estimates (0.04-0.4\,dex) but for a much larger sample. The spatial distribution of CTTS align with the location of protostars and molecular gas suggesting that they retain an imprint of the natal gas fragmentation process. Strong accretors are concentrated spatially, while weak accretors are more distributed. Our results do not support the sequential star forming processes suggested in the literature.    
\end{abstract}
 

\setlength{\parskip}{0.1 cm plus2mm minus2mm}
\setlength{\textfloatsep}{0.5 cm}  

\section{Introduction}

Classical T-Tauri stars (CTTS) are low-mass pre-main sequence (PMS) stars thought to accrete mass 
from a circumstellar disc via a stellar magnetosphere. Mass accretion rates ($\dot{M}_{\rmn{acc}}$) are thought to decrease with time as the circumstellar disc material is depleted. 
Simple viscous disc evolution predicts the observed trend rather well (\citealt{sici06, geert11, manara12}).
Observations also reveal that $\dot{M}_{\rmn{acc}}$ scales steeply to the square of the stellar mass 
$M_{*}$ (\citealt{muz03}; \citealt*{natta06}; \citealt{sici06, rig11, manara12, dol25}) in a variety of environments, however no such dependence is actually predicted by current 
theory (\citealt{hart06}). 
Observational $\dot{M}_{\rmn{acc}}$ estimates can provide an empirical base against which disc evolution 
models can be tested, and star formation histories of individual star-forming regions can be de-constructed (\citealt{geert11, venuti14}). Selection effects \citep{clark06} and intrinsic variability (\citealt{schol05, cost12, cost14}) may however affect the observed relations, and for these reasons large, uniform statistical samples are essential to constrain theory.  
     
Empirical $\dot{M}_{\rmn{acc}}$ values have been derived from intermediate-to-high resolution 
spectra (\citealt{muz03, natta06, sici06}), $U$-band (\citealt{rig11}) or H$\alpha$ photometry (\citealt{geert11,de10,spezzi12,manara12}). 
Statistical studies of $\dot{M}_{\rmn{acc}}$ have generally been restricted to relatively nearby 
star-forming regions, such that the effects of either massive stars or the variation of the star-forming environment on PMS evolution remain poorly constrained. 
In particular, whether photoevaporation from OB stars in the local environment 
affects $\dot{M}_{\rmn{acc}}$ remains an open question \citep{will11}. 
     
The advent of deep, high-resolution wide field narrow and multi-band surveys of the Galactic plane affords new opportunities to study star forming regions in greater details and to uncover significant new samples of CTTS in a variety of star-forming regions to address the problems raised in the preceding paragraphs. The VST Photometric H$\alpha$ survey of the southern  
Galactic plane and bulge (VPHAS+) is an $ugri$H$\alpha$ imaging survey reaching down to $r$\,$\sim$\,21\,mag \citep{drew14}. 
$\dot{M}_{\rmn{acc}}$ rates can potentially be estimated photometrically from either the $u$-band or H$\alpha$ excess 
with known detection limits. It thus becomes possible to construct a sample of 
$\dot{M}_{\rmn{acc}}$ rates that is relatively complete for PMS masses down to 0.5\,$M_{\odot}$ at 2\,kpc, assuming moderate extinction ($A_{V}$\,$\sim$\,2). 
This enables a uniform $\dot{M}_{\rmn{acc}}$ data-set in a variety of notable star-forming regions, including 
the Carina Nebula, NGC\,6611, and NGC\,3603. 
Such work has already begun by \cite{vink08} in the Cygnus OB\,2 massive star forming region, \cite{geert11} in the IC\,1396 star-forming region, and \cite{geert13} in the open cluster NGC\,2264 using data from the counterpart northern Galactic plane survey IPHAS \citep{drew05}.

\begin{figure*}        
\center
\includegraphics[width=170mm, height=100mm]{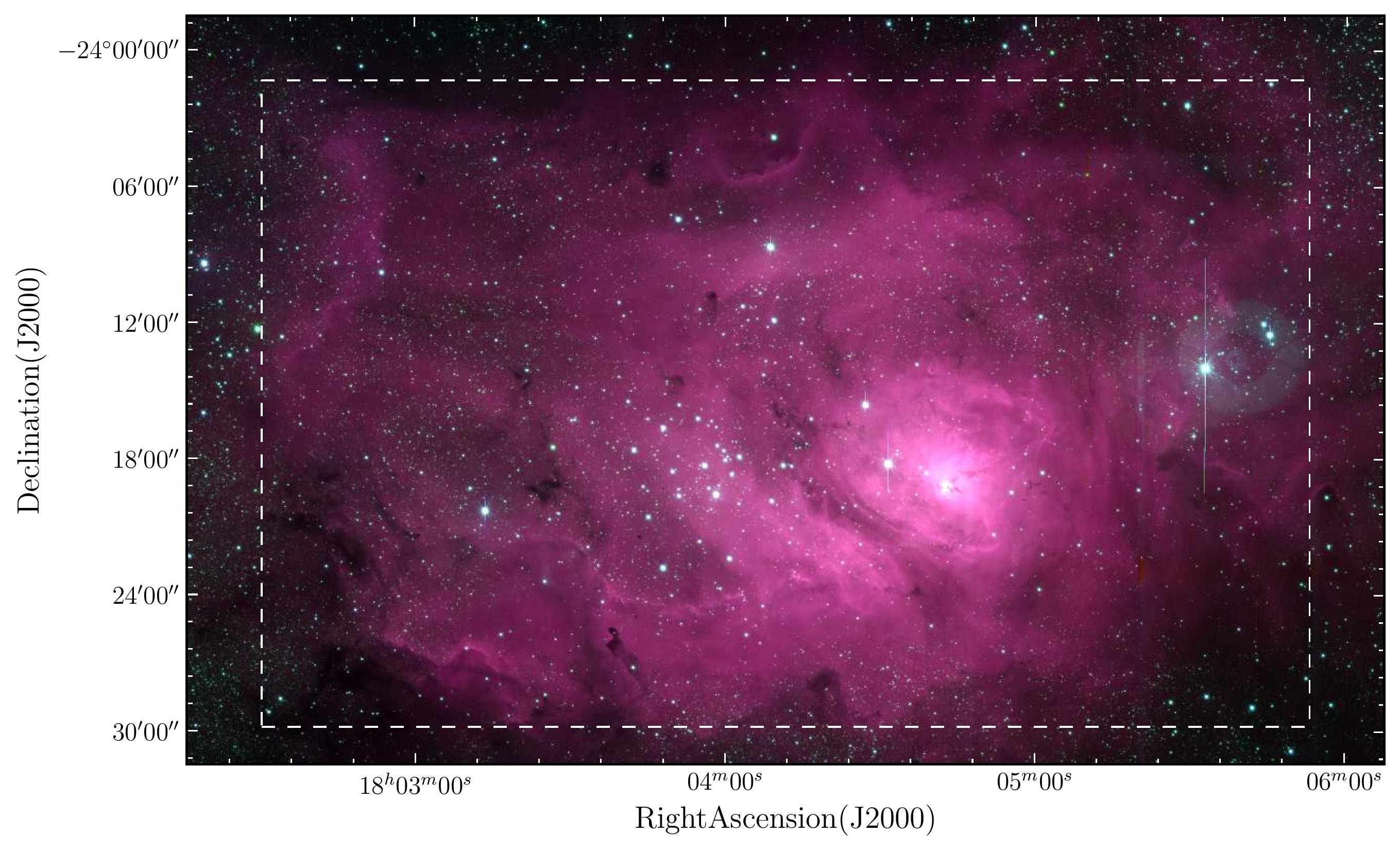}
\caption{VPHAS+ $ri$H$\alpha$ three-colour image of the Lagoon Nebula. North is up and east is to the left. The H{\scriptsize{II}} region Lagoon Nebula (or M8) is marked by the box having dimensions 50$\arcmin$\,$\times$\,30$\arcmin$.}
\label{fig:im}
\end{figure*}  

In this paper, we aim to measure $\dot{M}_{\rmn{acc}}$ for a significant fraction of CTTS candidates 
in the Lagoon Nebula (M8) identified using VPHAS+ survey photometry. M8 is a southern H{\scriptsize{II}} region located in the Sagittarius-Carina arm. It comprises the embedded open cluster NGC\,6530 located to the east of the central dust lane, and the optically obscured Hourglass nebula cluster to its west (see Fig. 1). It contains at least sixty known PMS members (\citealt*{sung00}; \citealt{pris05}; \citealt*{arias07}; \citealt{tothill08}), but with 
very few $\dot{M}_{\rmn{acc}}$ estimates in literature \citep*{gall12}. The median age of PMS members in the region has been estimated $\sim$\,1\,Myr using colour-magnitude diagrams (CMD; \citealt{sung00, pris05, mayne07, arias07}), concurrent with both the dynamical age (\citealt*{chen07, van72}) and the main-sequence position of the confirmed O-type members (\citealt{tothill08}). It is comparatively younger than most regions for which a statistically significant number of $\dot{M}_{\rmn{acc}}$ estimates are known (e.g. see Fig. 8 in \citealt{spezzi12}) to enable a meaningful statistical analysis. However, it has been suggested that a previous burst of star formation occurred around $\sim$\,10\,Myr ago by \cite{lada76} and \cite{ancker}. \cite{lightfoot84, dami04, pris05, arias07, pris12} have reported sequential age gradients of the PMS members in region.   

The mean reddening $E(B-V)$ towards NGC\,6530 measured using {\it{UBV}} colour diagrams \citep{sung00}, SED-fitting \citep{ancker, arias06} and isochrone fitting (\citealt{pris05, mayne08}) is 0.35\,mag. The highest measured $E(B-V)$ in these studies is $\sim$\,0.5\,mag for a few stars, and the foreground extinction $\sim$ 0.2\,mag. Distance ($d$) determinations to NGC\,6530 using isochrone fitting provide $d$\,=\,1250$\pm$50\,pc (\citealt{pris05, arias07, mayne08}). Previous measurements indicate $d$ as high as 1800\,pc (\citealt{ancker, sung00}). This discrepancy is thought to arise because larger determinations rely on fitting only O and early B type stars, rather than the complete blue edge of the distribution of stars in the upper main sequence on the CMD. \cite{pris05} argue that any background stars lying behind the nebula must be significantly reddened, moving them away from the main sequence. Based on the above argument, \cite{pris05} suggest that the blue edge of the distribution of stars in the CMD more accurately defines the zero-age main sequence at the distance to the Lagoon Nebula.

Most authors \citep{neckel81, sung00, ancker, arias02, arias06} have found the reddening to follow a normal reddening law, $R_{V}$\,$\approx$\,3.1, but found that H\,36, an O7\,V star located in the Hourglass nebula with $R_{V}$\,ranging from 4\,-\,8 \citep{john67,sung00}. This anomaly is explained by the presence of larger silicate grains surrounding H\,36, suggesting selective evaporation in its immediate environment leading to an anomalous $R_{V}$ in the close vicinity \citep{hec82}. Sung et al. (2000) used the $E(B-V)/E(V-I)$ colour-excess ratio and found a $R_{V}$ range between 3.0\,-\,3.6 for OB stars in the Lagoon Nebula with a mean of 3.1, but higher values for PMS stars which were effected by blue excess emission, and H36. We note that Prisinzano et al. (2012) used the longer baseline afforded by the $E(B-V)/E(V-K)$ ratio to find $R_{V}$\,$\sim$\,5, inconsistent with standard reddening. However, many of the stars used to derive this reddening law were PMS stars that are most likely affected by excess blue and infrared emission. Prisinzano et al. (2012) note that because of this, they were unable to constrain $R_{V}$ with precision. We note that an anomalous reddening law of $R_{V}$\,=\,5 leads to a difference $A_{r}$ of $\approx$\,0.5, which is approximately the same as the spread in $E(B-V)$ values.   

At the mean reddening and distance to the Lagoon Nebula 0.3\,$M_{\odot}$-\,2\,$M_{\odot}$ PMS stars span a magnitude range 13\,$<$\,$r$\,$<$\,21, making VPHAS+ photometry suited to estimate their $\dot{M}_{\rmn{acc}}$. The four confirmed O-type members plus some sixty B-type stars \citep{tothill08} represent a step toward increasing the massive star content compared to similar works using IPHAS in the Northern Galactic plane (\citealt{geert11, geert13}).

In Section 2, VPHAS$+$ observations and source selection criteria are detailed. 
In Section 3, we identify CTTS candidates. We measure the accretion rates and stellar properties of the CTTS candidates in Sect.\,4. The distribution of $\dot{M}_{\rmn{acc}}$ with respect to stellar mass is discussed in Section 5. The spatial distribution of CTTS, and their properties is presented in Section 6. Finally, our conclusions are presented in Section 7. 
 
\section{Data}
VPHAS+ imaging is obtained by the OmegaCAM CCD imager mounted on the 2.6\,metre VLT Survey Telescope (VST) on Cerro Paranal, Chile. The imager captures a $1^{\circ}$ square field of view at a resolution of 0.21$\arcsec$\,pixel$^{-1}$. Each pointing is supplemented by at least one offset exposure to minimise CCD gaps. Imaging is carried out through broadband {\it ugri} filters and a purpose-built H$\alpha$ filter. The central wavelength and bandpass of the H$\alpha$ filter are 6588 and 107\,\AA~respectively. Exposure times are 150, 30, 25, 25, and 120\emph{s} respectively. VPHAS+ reaches a 5$\sigma$ depth at H$\alpha$\,=\,20.5-21.0\,mag and $g$\,=\,22.2\,-\,22.7\,mag. Practical constraints have meant that the blue ($ug$) and red ($ri$H$\alpha$) observations are carried out separately. An additional $r$ observation is carried out with every blue observation to serve as a linking reference. 

Observations used in this paper are detailed in Table~\ref{tab:obs}. Data from the survey are pipeline processed at the Cambridge Astronomical Survey Unit (CASU), where nightly photometric calibrations are performed. Astrometry of sources is refined by matching with the 2MASS catalogue \citep{cutri03}. 

Overlaps between the observations were used to bring the red and blue data onto a common internal scale. The calibrations of band-merged catalogues were checked and refined using external $gri$ data from the APASS all-sky catalogue (http://www.aavso.org/apass). We cross-matched bright sources (13\,$<$\,mag\,$<$\,15) morphologically classified as stellar in the VPHAS+ bands, and having random photometric error in each catalogue less than 0.1\,mag. Around 200 sources were selected for comparison in each band. The median magnitude offsets found were $\Delta r$(red)\,=\,$-$0.027, $\Delta r$(blue)\,=\,$-$0.029, $\Delta i$\,=\,$-$0.0106, $\Delta g$\,=\,0.044. $\Delta u$\,=\,$-$0.23\,mag was determined following the procedure described in Drew et al. (2014). 

\begin{table}
\caption{Log of observations}
\label{tab:obs}
\centering
\resizebox{\linewidth}{!}{
\begin{tabular}{|l|l|l|l|l|}
\hline
  \multicolumn{1}{|c|}{Right Ascension} &
  \multicolumn{1}{c|}{Declination} &
  \multicolumn{1}{c|}{Filters} &
  \multicolumn{1}{c|}{Obs. Date} &
  \multicolumn{1}{c|}{Seeing} \\
\hline
  18$^{h}$02$^{m}$06.32$^{s}$ & $-$24$\degr$11$\arcmin$36.8$\arcsec$ & $ri$H$\alpha$ & 2012 June 10 & 0.6$\arcsec$\\
  18$^{h}$01$^{m}$23.47$^{s}$ & $-$24$\degr$00$\arcmin$36.0$\arcsec$ & $ri$H$\alpha$ & 2012 June 10 & 0.6$\arcsec$\\
  18$^{h}$06$^{m}$18.95$^{s}$ & $-$24$\degr$11$\arcmin$36.8$\arcsec$ & $ri$H$\alpha$ & 2012 June 27 & 0.7$\arcsec$\\
  18$^{h}$05$^{m}$36.10$^{s}$ & $-$24$\degr$00$\arcmin$36.0$\arcsec$ & $ri$H$\alpha$ & 2012 June 27 & 0.7$\arcsec$\\
  18$^{h}$02$^{m}$06.32$^{s}$ & $-$24$\degr$11$\arcmin$36.8$\arcsec$ & $ugr$ & 2012 Aug 14 & 0.8$\arcsec$\\
  18$^{h}$01$^{m}$23.47$^{s}$ & $-$24$\degr$00$\arcmin$36.0$\arcsec$ & $ugr$ & 2012 Aug 14 & 0.8$\arcsec$\\
  18$^{h}$06$^{m}$18.95$^{s}$ & $-$24$\degr$11$\arcmin$36.8$\arcsec$ & $ugr$ & 2012 Aug 15 & 0.9$\arcsec$\\
  18$^{h}$05$^{m}$36.10$^{s}$ & $-$24$\degr$00$\arcmin$36.0$\arcsec$ & $ugr$ & 2012 Aug 15 & 0.9$\arcsec$\\
\hline\end{tabular}}
\end{table}  
 
\subsection{Source selection}

The area of study is centred on RA 18$^{h}$04$^{m}$12$^{s}$, Dec $-$24$\degr$18$\arcmin$36$\arcsec$ (J2000) and covers 50$\arcmin$\,$\times$\,30$\arcmin$ (Fig. 1). This region encloses the Lagoon Nebula (Tothill et al. 2008). We apply the following criteria to select sources: 

\begin{enumerate}
\item $r$\,$>$\,13\,mag in both the red and blue filter sets to avoid saturated sources;
\item random photometric error must be less than 0.1\,mag in $g$,$r$(red and blue),$i$ and H$\alpha$ to keep photometric and propagated uncertainties small;
\item morphologically classified as stellar in $g$ and $i$ and either stellar or star-like in {\it{r}} (see Table 7 in \citealt{gonz08}) to avoid extended objects. 
\item morphologically classified as either stellar, star-like or extended in H$\alpha$ and $u$-band photometry. The reasoning behind these criterion are given in the following paragraph. 
\item average pixel confidence value greater than 90 and 95 per\,cent in broadband $ugri$ and H$\alpha$ respectively (see Drew et al. 2014). This is especially important in H$\alpha$ to ensure the photometry remains unaffected by extra vignetting by the T bars separating the H$\alpha$ filter segments. 
\end{enumerate}

\begin{figure}
\center
\includegraphics[width=72mm, height=78mm]{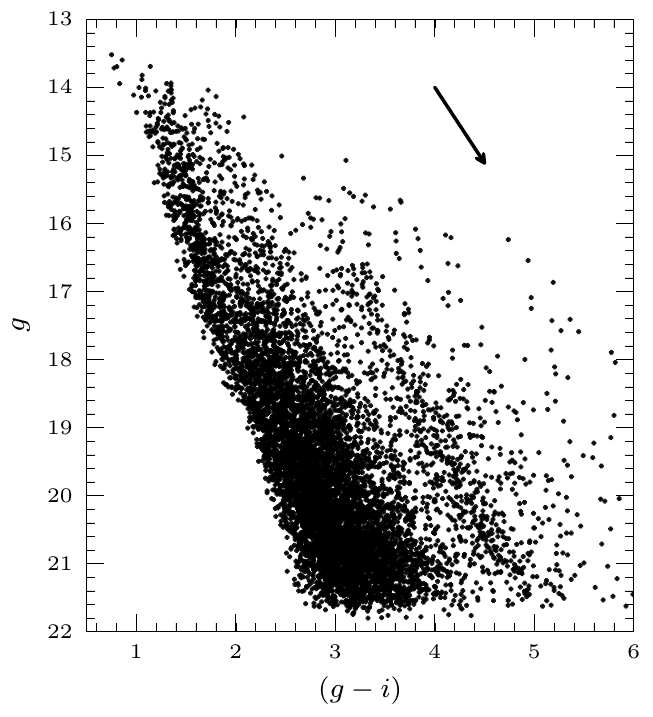}
\caption{The $g$ versus ($g-i$) CMD. The diagram includes all point sources meeting our selection criteria in $g,i$ located in the Lagoon Nebula (see Fig.1). A reddening vector for $A_{V}$\,=\,1 is shown.}
\label{fig:cmdgi}
\end{figure}

To filter for background variations, the CASU {\it{nebuliser}} routine is performed on the VPHAS+ images as part of the data processing pipeline. The background is modelled using a sliding median filter over a scale of 15$\arcsec$. Although rapidly spatially varying nebulosity on shorter spatial scales can lead to imprecise sky subtractions. This is especially pertinent when considering the variation of nebulosity on small scales as is the case in the regions surrounding the NGC\,6530 cluster and the Hourglass nebula (see Fig. 1). Imperfect sky subtractions or imperfect flat-fielding in such cases can affect the point spread function of sources especially in H$\alpha$ and $u$-band photometry respectively. In such cases, the final morphological classification of H$\alpha$ or $u$-band sources assigned by the pipeline may be +1 (extended), reflecting the resulting mild degradation of the point spread function. So as not to reject too many detections of sources that are otherwise clearly point-like, we relax the morphological constraints and accept `extended' H$\alpha$ and $u$ measurements. Instead, we inspect the images of each selected H$\alpha$ excess emission line source by eye. An unsharp mask filter was applied to the reduced and calibrated single-band mosaics to suppress large scale low frequency features such as diffuse background, and highlight the high frequency sharp nebulous features in particular those varying on scales less than the median background filter. On this, the standard photometric aperture (of 1$\arcsec$), and the associated background annulus of each source were marked. Any sources with non-uniform filamentary structures or nebulosity within the background annulus were excluded from further study. It is noted that previous similar work within the Lagoon Nebula has faced similar difficulties, including the slit spectroscopic work of Arias et al. (2007) where the authors were unable to subtract accurately the background emission in select cases, and in the fibre spectroscopic work of Prisinzano et al. (2007; 2012) where the authors were unable to measure accurately H$\alpha$ equivalent widths of stellar sources due to difficulty in isolating the background nebulosity from stellar emission.  In particular, the bright and complex nebulosity around the Hourglass nebula (a 2.6$\arcmin$\,$\times$\,2.6$\arcmin$ region centred on RA 18$^{h}$03$^{m}$41$^{s}$, Dec $-$24$\degr$22$\arcmin$18$\arcsec$) varies on scales less than a few {\it{arcsecs}}, and makes it extremely difficult to isolate point-source H$\alpha$ emission. Hence, all H$\alpha$ sources within this region were excluded. Based on the single-slit observations of Arias et al. (2007), we estimate that up to eight known accreting PMS visible in the optical may have been omitted because of removing this region. 

In the resulting sample, 12,384 sources have $ri$H$\alpha$ photometry. A subset of 11,060 have $g$-band photometry and 4,864 have $u$-band photometry. The selected sources are shown in the ($g-i$) versus $i$ CMD in Fig.~\ref{fig:cmdgi}. The saturation limit of the VPHAS+ photometry ($g$\,$\sim$\,14\,mag) means that an upper main sequence containing the bright OB stars of the Lagoon are saturated is not visible. The lack of faint background stars first remarked by Prisinzano et al. (2005) is also seen, and is interpreted as a blue envelope due to the molecular cloud preventing the detection of background stars at visible magnitudes (see Section 1). We also note a sequence of reddened late-type dwarf stars in the range (3.2\,$<$\,($g-i$)\,$<$\,5) and (17\,$<$\,$g$\,$<$\,22) also found in the CMD of Sung et al. (2000) and Prisinzano et al. (2005).

\section{Identifying Classical T\,Tauri candidates using photometric H$\alpha$ equivalent width} 

The ($r-$H$\alpha$) colour is a measure of H$\alpha$ line strength relative to the $r$-band photospheric continuum. Main-sequence stars do not have H$\alpha$ in emission. Modelling their ($r-$H$\alpha$) colour at each spectral type allows for a template against which any colour excess due to H$\alpha$ emission can be measured from the observed ($r-$H$\alpha$) colour. The ($r-$H$\alpha$) colour excess (defined as ($r-$H$\alpha$)$_{\rmn{excess}}$\,=\,($r-$H$\alpha$)$_{\rmn{observed}}$$-$($r-$H$\alpha$)$_{\rmn{model}}$) can be used to compute the H$\alpha$ equivalent width (EW$_{{\rmn{H}}\alpha}$) (see \citealt{de10}).  

We use the ($r-i$) colour as a  proxy for the spectral type. This is a reasonable assumption for late spectral types (K\,-\,M6) which form the majority of our observed sample given the 1.5\,mag spread in the ($r-i$) colour across this spectral range, and a mean ($r-i$) colour error due to the combination of random photometric noise and extinction uncertainty $\lesssim$\,0.15\,mag. This results in the average error on the estimated spectral type caused by using the ($r-i$) colour as a proxy for spectral classification to be between 2-3 spectral subclasses. To calculate the main-sequence colours, \cite{pick98} spectra for O5-M6 spectral types were convolved with the appropriate filter band passes and CCD response. 
 
\begin{figure}
\center 
\includegraphics[width=86mm, height=58mm]{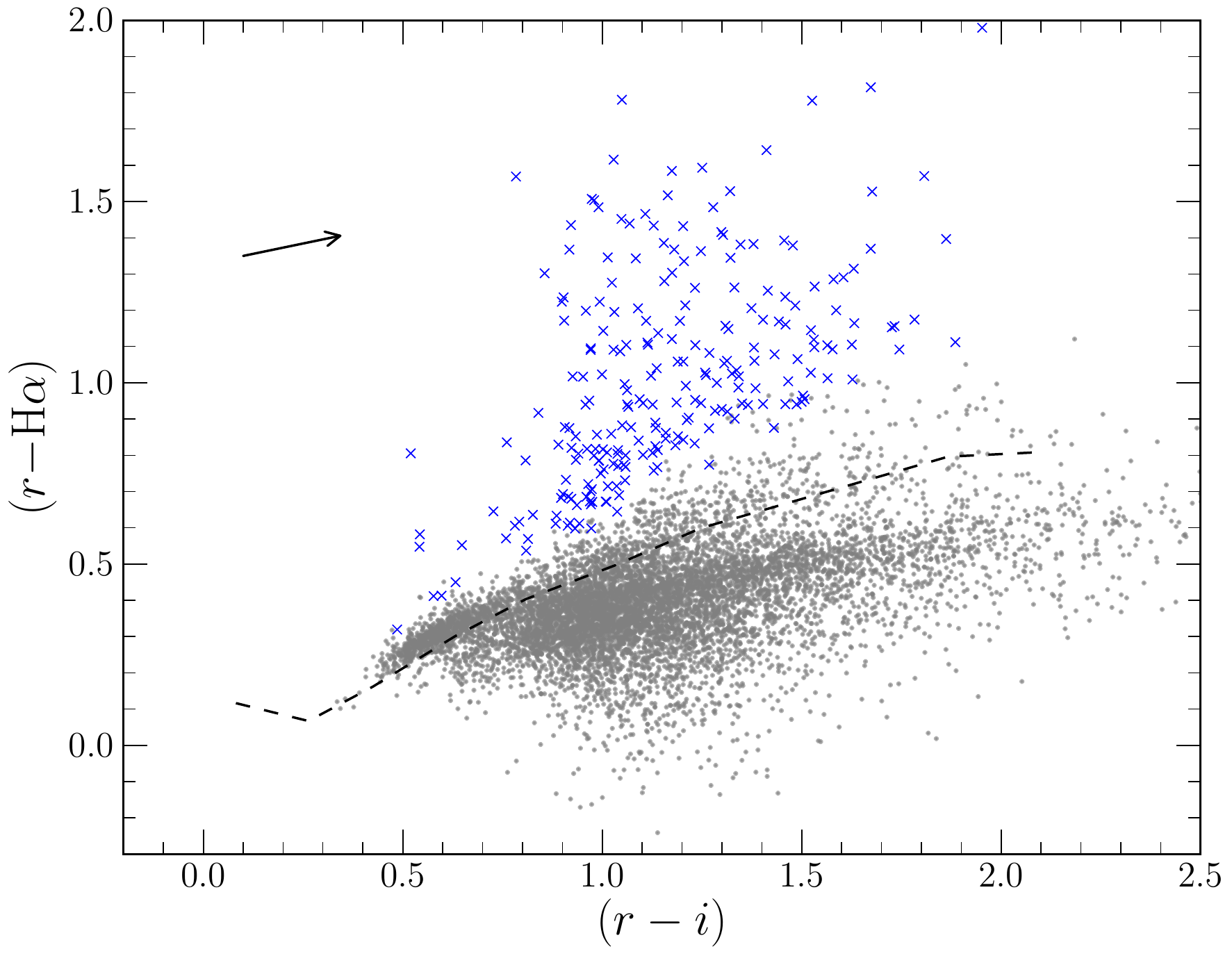}
\center
\caption{($r-$H$\alpha$) vs. ($r-i$) diagram. The dashed line is the interpolated model track reddened by $A_{V}$\,=\,1. Grey dots are stars meeting our selection criterion, and blue crosses are stars meeting our selection criterion that are identified as candidate CTTS based on their EW$_{\rmn{H}\alpha}$. A reddening vector for $A_{V}$\,=\,1 is shown.}
\label{fig:rha} 
\end{figure}

In Fig.~\ref{fig:rha} the ($r-$H$\alpha$) vs. ($r-i$) diagram is plotted. The dashed line represents the interpolated model colours reddened by the mean reddening $E(B-V)$\,=\,0.35 (see Sec. 1). At a given ($r-i$), the ($r-$H$\alpha$)$_{\rmn{excess}}$ for each star is computed. The EW$_{\rmn{H}\alpha}$ is given by:
  \begin{equation}
    \ \rmn{EW}_{\rmn{H}\alpha}= \rmn{W}\times[1-10^{0.4\times(r-\rmn{H}\alpha)_{\rmn{excess}}}]\
  \end{equation}
as shown by De Marchi et al. (2010). W is the rectangular bandwidth of the H$\alpha$ filter. The photometric EW$_{\rmn{H}\alpha}$ for all stars having $ri$H$\alpha$ magnitudes in our sample are measured. We validate our measured EW$_{\rmn{H}\alpha}$ against spectroscopic measurements in Section 3.1. 

\begin{figure} 
\center
\includegraphics[width=86mm, height=58mm]{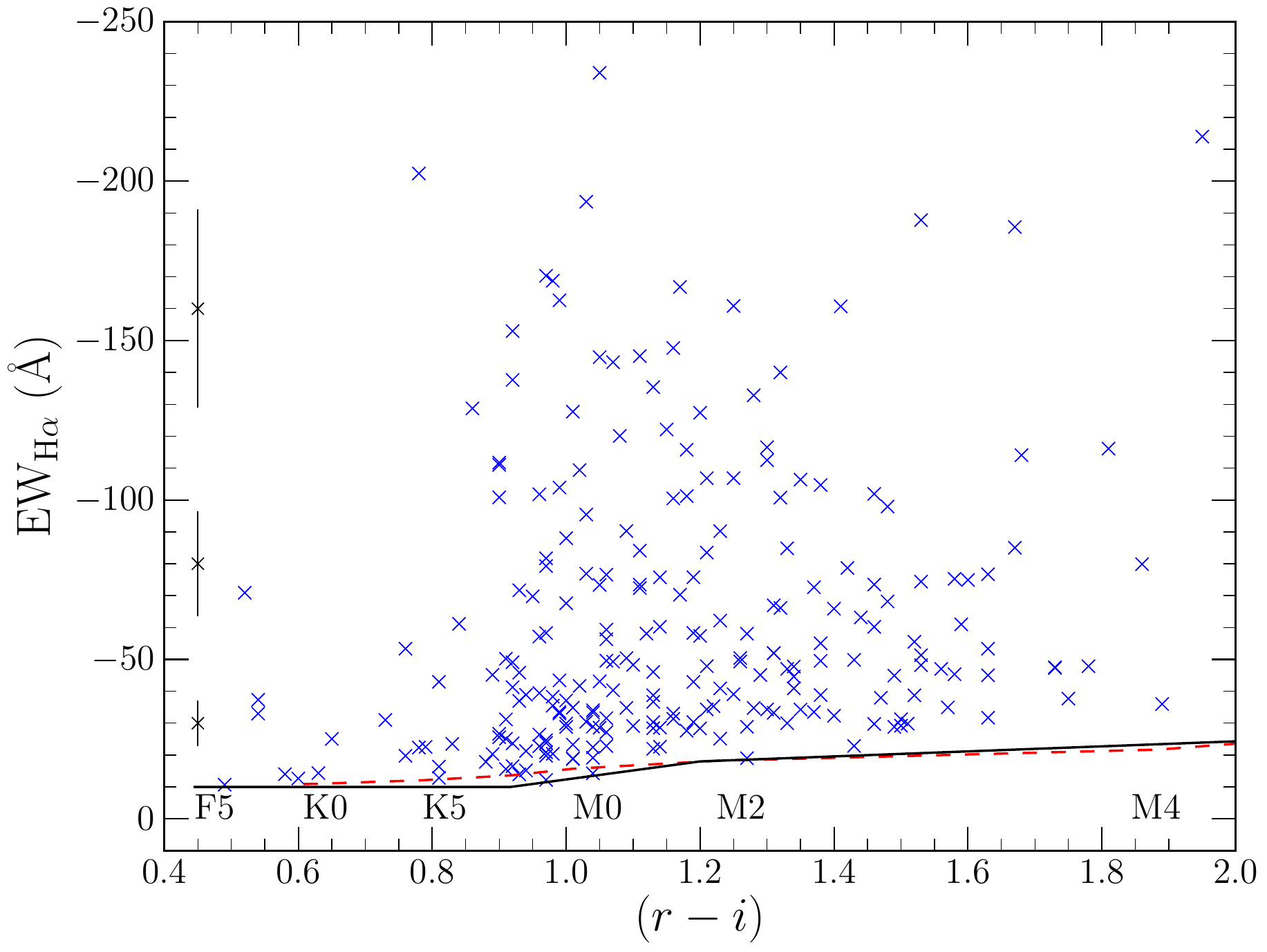}
\caption{The results of the spectral type-EW$_{\rmn{H}\alpha}$ selection criteria. Ordinate shows the $(r-i)$ colour with the corresponding spectral type labelled, while the abscissa shows the EW$_{\rmn{H}\alpha}$ of selected CTTS. The median error corresponding to a given EW$_{\rmn{H}\alpha}$ is shown at ($r-i$)\,=\,0.45. The solid line is the empirical selection criterion of White \& Basri (2003), and the red dashed line is the spectral type-EW$_{\rmn{H}\alpha}$ criteria of Barrado y Navascu{\'e}s and Mart{\'i}n (2003) is shown for comparison.}
\label{fig:riEW}
\end{figure}

We homogeneously select CTTS from the sample of stars having measured EW$_{\rmn{H}\alpha}$ using the spectral type-EW$_{\rmn{H}\alpha}$ criteria of \cite{white03}. We note that the median EW$_{\rmn{H}\alpha}$ errors due to photometric and reddening uncertainties vary between $\sim$\,5\,-\,7\,\AA~depending on spectral type, so we empirically adjust the selection criteria of White \& Basri (2003) by the same amount. We consider CTTS as those having spectral type F5-K5 and EW$_{\rmn{H}\alpha}$\,$<$\,$-10$\,\AA, K5-K7.5 and EW$_{\rmn{H}\alpha}$\,$<$\,$-12$\,\AA~ and M2.5-M6 and EW$_{\rmn{H}\alpha}$\,$<$\,$-25$\,\AA. The results of our selection procedure are shown in Fig.~\ref{fig:riEW}. The selection criteria used are designed to exclude chromospherically active late type stars, whose maximum measured EW$_{\rmn{H}\alpha}$ ranges from $-$11\,\AA~for a M2V spectral type to $-$24\,\AA~for an M6V spectral type \citep{barr03}. We select 235 candidate CTTS on this basis, which constitute the sample discussed in the rest of this paper. The photometric properties of the selected CTTS are given in Table 2. 

\begin{table*}
\caption{VPHAS+ photometry of candidate CTTS in the Lagoon Nebula. The full table is available in the online version of this journal.}
\begin{tabular}{|l|l|l|l|l|l|l|l|}
\hline
  \multicolumn{1}{c}{JNAME} &
  \multicolumn{1}{c}{RA} &
  \multicolumn{1}{c}{Dec} &
  \multicolumn{1}{c}{$r$} &
  \multicolumn{1}{c}{($r-i$)} &
  \multicolumn{1}{c}{($r-$H$\alpha$)} &
  \multicolumn{1}{c}{($g-r$)} &
  \multicolumn{1}{c}{($u-g$)} \\
  \multicolumn{1}{c}{} &
  \multicolumn{1}{c}{(J2000)} &
  \multicolumn{1}{c}{(J2000)} &
  \multicolumn{1}{c}{(mag)} &
  \multicolumn{1}{c}{} &
  \multicolumn{1}{c}{} &
  \multicolumn{1}{c}{} &
  \multicolumn{1}{c}{} \\
\hline
  18023206$-$2415320 & 270.63358 & $-$24.25888 & 14.83$\pm$0.001 & 0.63$\pm$0.002 & 0.45$\pm$0.002 &  & \\
  18023355$-$2414548 & 270.63977 & $-$24.24856 & 16.40$\pm$0.003 & 1.01$\pm$0.004 & 0.67$\pm$0.006 &  & \\
  18023368$-$2418022 & 270.64032 & $-$24.30061 & 19.96$\pm$0.049 & 1.95$\pm$0.051 & 1.98$\pm$0.067 &  & \\
  18023560$-$2413024 & 270.64832 & $-$24.21733 & 18.29$\pm$0.012 & 1.44$\pm$0.014 & 1.17$\pm$0.023 & 1.89$\pm$0.029 & 0.16$\pm$0.065\\
  18023730$-$2416242 & 270.65543 & $-$24.27338 & 16.51$\pm$0.003 & 1.13$\pm$0.004 & 0.81$\pm$0.006 &  & \\
  18023836$-$2419313 & 270.65985 & $-$24.32536 & 18.52$\pm$0.014 & 1.13$\pm$0.018 & 0.88$\pm$0.032 & 1.03$\pm$0.019 & $-$0.21$\pm$0.026\\
  18023897$-$2414274 & 270.66238 & $-$24.24094 & 17.81$\pm$0.009 & 0.90$\pm$0.012 & 1.24$\pm$0.014 &  & \\
  18023972$-$2419310 & 270.6655 & $-$24.32527 & 18.45$\pm$0.014 & 1.31$\pm$0.016 & 0.92$\pm$0.027 & 1.76$\pm$0.036 & 0.40$\pm$0.104\\
  18024000$-$2419346 & 270.66666 & $-$24.32627 & 17.67$\pm$0.007 & 1.16$\pm$0.009 & 0.85$\pm$0.014 & 1.63$\pm$0.015 & 0.36$\pm$0.039\\
  18024097$-$2412163 & 270.67072 & $-$24.20452 & 16.59$\pm$0.004 & 0.96$\pm$0.005 & 0.82$\pm$0.009 & 1.43$\pm$0.006 & 0.25$\pm$0.013\\
\hline\\
\end{tabular}
\end{table*}

\subsection{Comparison of photometric and spectroscopic EW$_{\rmn{H}\alpha}$}

To test the accuracy of our photometrically determined EW$_{\rmn{H}\alpha}$, we compare with spectroscopically measured EW$_{\rmn{H}\alpha}$ from \cite{arias07}. Arias et al. (2007) obtained intermediate resolution spectra covering the H$\alpha$ line from the Magellan telescope (Las Campanas Observatory, Chile) of 46 stars in the Lagoon Nebula. Stars were selected for observation on the basis of either H$\alpha$ excess in $RI$H$\alpha$ photometry from Sung et al. (2000), disc excess in near-infrared {\it{JHK}}s photometry, or extended or knotty morphological appearance in {\it{Hubble}} images. Out of the 46 stars observed by Arias et al. (2007), 37 were identified as H$\alpha$ emission line stars. Twenty seven stars have VPHAS+ $ri$H$\alpha$ photometry. None of the ten stars classified as H$\alpha$ emission-line stars that are not detected in VPHAS+ H$\alpha$ photometry have literature H$\alpha$ photometry in ground-based imaging from Sung et al. (2000). Of the ten, six were selected on the basis of their morphological appearance in {\it{Hubble}} imaging by Arias et al. (2007) and cannot be resolved to point sources in VPHAS+ H$\alpha$ imaging (ID. 3-8 inclusive in Table 2 and 3 of Arias et al. 2007; all located in the Hourglass nebula), or ground-based H$\alpha$ imaging in Sung et al. (2000). Two are surrounded by intense nebulosity and cannot be resolved as point sources. Two sources were selected for observations by \cite{arias07} using near-infrared photometry and do not have resolvable optical counterparts (ID. 1, 9, 10, 11, in Table 2 and 3 of Arias et al. 2007; all located in the Hourglass Nebula). A further two stars having $ri$H$\alpha$ photometry are saturated, and their photometric EW$_{\rmn{H}\alpha}$ are possibly incorrect (ID. 15, 45 in Table 2-3 of Arias et al. 2007). 

\begin{table*}
\caption{Comparison of detected H$\alpha$ emission line stars with spectroscopic EW$_{\rmn{H}\alpha}$ from Arias et al. (2007; A07), and archive data from Sung et al. (2000; S00).}
\begin{tabular}{|l|l|l|l|l|l|l|l|l|l|l|}
\hline
  \multicolumn{1}{|c|}{ID} &
  \multicolumn{1}{c|}{$r$} &
  \multicolumn{1}{c|}{$r-$H$\alpha$} &  
  \multicolumn{1}{c|}{$V$} &
  \multicolumn{1}{c|}{$R-$H$\alpha$} &  
  \multicolumn{1}{c|}{Sp. EW$_{\rmn{H}\alpha}$} &
  \multicolumn{1}{c|}{EW$_{\rmn{H}\alpha}$} &
  \multicolumn{1}{c|}{CTTS} &  
  \multicolumn{1}{c|}{Comment} \\
  \multicolumn{1}{|c|}{A07/S00} &
  \multicolumn{1}{c|}{(mag)} &
  \multicolumn{1}{c|}{} &  
  \multicolumn{1}{c|}{(mag)} &
  \multicolumn{1}{c|}{} &  
  \multicolumn{1}{c|}{(\AA)} &
  \multicolumn{1}{c|}{(\AA)} &
  \multicolumn{1}{c|}{} &
  \multicolumn{1}{c|}{} \\ 
\hline
12/174 & 16.27 & 1.05 & 16.90 &   & $-$15.1\,$\pm$\,0.1 & $-$73.5\,$\pm$\,12 & N & No lit. H$\alpha$, near HG \\ 
  &   &   &  &  &  &  &  & Does not meet photometric criteria\\
13/184$^1$ & 15.66 & 1.44 & 16.53 & $-$2.88 & $-$126\,$\pm$\,9 & $-$138.3\,$\pm$\,30 & Y &  \\
15/240$^1$ & 10.63 & 0.54 & 11.72 & $-$4.23 & $-$24.1\,$\pm$\,0.4 & $-$46.8\,$\pm$\,2 & N & Saturated, HAeBe star \\  
17/292$^1$ & 16.25 & 0.95 & 16.84 & $-$4.06 & $-$52\,$\pm$\,3 & $-$58.3\,$\pm$\,11 & Y &  \\ 
18/388 & 15.07 & 0.57 & 16.70 &   & $-$3.8\,$\pm$\,0.2 & $-$12.8\,$\pm$\,2 & Y & No lit. H$\alpha$ \\ 
19/390$^1$ & 15.22 & 0.54 & 15.58 & $-$3.75 & $-$16\,$\pm$\,0.6 & $-$33.5\,$\pm$\,5 & Y & \\ 
21/418$^1$ & 14.81 & 0.6 & 16.5 & $-$3.98 & $-$5.6\,$\pm$\,0.03 & $-$13.4\,$\pm$\,1 & Y & \\ 
22/422 & 14.78 & 0.82 & 15.53 &  & $-$27.8\,$\pm$\,0.5 & $-$35.8\,$\pm$\,3 & Y & No lit. H$\alpha$ \\ 
24/440$^1$ & 14.1 & 0.44 & 15.99 & $-$4.14 & $-$0.65\,$\pm$\,0.03 & $-$0.75\,$\pm$\,1 & N & Does not meet EW$_{\rmn{H}\alpha}$--Sp. type criteria \\ 
25$^2$ & 16.47 & 0.46 &  &  & $-$8.1\,$\pm$\,0.3 &  & N & No lit. H$\alpha$, no H$\alpha$ emission detected \\ 
26/482 & 13.95 & 0.50 & 15.45 & $-$4.22 & $-$9.3\,$\pm$\,0.2 & $-$8.6\,$\pm$\,1 & N & Does not meet EW$_{\rmn{H}\alpha}$--Sp. type criteria\\ 
27/486$^1$ & 16.79 & 1.14 & 16.9 & $-$3.26 & $-$75\,$\pm$\,4 & $-$82\,$\pm$\,17 & Y &  \\ 
29/495$^1$ & 15.89 & 0.72 & 16.73 & $-$3.80 & $-$29\,$\pm$\,1 & $-$22.5\,$\pm$\,4 & Y & \\ 
30/508$^1$ & 16.44 & 1.54 & 16.73 & $-$3.73 & $-$30.3\,$\pm$\,0.7 & $-$155.7\,$\pm$\,33 & Y & Marked as eruptive variable star \\ 
  &   &   &  &  &  &  &  & in Kukarkin \& Kholopov (1982). \\
31/531 & 16.29 & 0.59 & 16.03 & $-$4.18 & $-$14.96\,$\pm$\,0.04 & $-$13.10\,$\pm$\,3 & Y &  \\ 
32/540$^1$ & 15.56 & 0.61 & 16.17 & $-$4.16 & $-$6.7\,$\pm$\,0.4 & $-$10.1\,$\pm$\,2.5 & Y & \\ 
33/547$^1$ & 14.33 & 0.6 & 14.93 & $-$4.16 & $-$2.26\,$\pm$\,0.07 & $-$17.4\,$\pm$\,2 & Y &  \\ 
34/552$^1$ & 14.55 & 0.71 & 15.39 & $-$3.80 & $-$12.8\,$\pm$\,0.3 & $-$49.7\,$\pm$\,7 & Y &  \\  
35/666$^1$ & 15.68 & 0.68 & 16.53 & $-$3.90 & $-$35.4\,$\pm$\,0.3 & $-$20.5\,$\pm$\,2 & Y &  \\ 
36/681 & 13.37 & 0.39 & 15.0 & $-$4.59 & $-$0.7\,$\pm$\,0.07 &  & N & No H$\alpha$ emission detected \\ 
40/726$^1$ & 16.07 & 0.8 & 16.88 & $-$3.91 & $-$48\,$\pm$\,2 & $-$45.9\,$\pm$\,3 & Y &  \\ 
41$^2$ & 13.20 & 0.35 &  &  & $-$3.18\,$\pm$\,0.06 & $-$2.09\,$\pm$\,1 & N &  No lit. H$\alpha$ \\ 
  &   &   &  &  &  &  &  & Does not meet EW$_{\rmn{H}\alpha}$-Sp. type criteria \\
42/782$^1$ & 15.47 & 0.87 & 16.25 & $-$3.55 & $-$54.0\,$\pm$\,2 & $-$50\,$\pm$\,4 & Y & \\ 
43/810$^1$ & 16.29 & 2 & 16.25 & $-$3.96 & $-$20.8\,$\pm$\,0.3 & $-$22.3\,$\pm$\,2 & Y & \\ 
44/862$^1$ & 15.34 & 0.96 & 16.90 & $-$3.26 & $-$22.3\,$\pm$\,0.4 & $-$41.8\,$\pm$\,4 & N & Does not meet photometric criteria\\ 
  &   &   &  &  &  &  &  & Located in M8\,E rim \\
    &   &   &  &  &  &  &  & Outlier in EW$_{\rmn{H}\alpha}$ vs. ($R-$H$\alpha$)  \\ 
  &   &   &  &  &  &  &  & diagram in Arias et al. (2007; Fig. 5)  \\   
45/879$^1$ & 10.74 & 0.93 & 11.94 & $-$3.15 & $-$80\,$\pm$\,1 & $-$101.8\,$\pm$\,10 & N & Saturated, HAeBe star\\ 
46$^2$ & 15.07 & 0.54 &  &  & $-$2.85\,$\pm$\,0.2 & $-$19.1\,$\pm$\,3 & N & No lit. H$\alpha$ \\
  &   &   &  &  &  &  &  & Does not meet photometric criteria\\ 
\hline
\end{tabular}\\
{\footnotesize{$^1$Object selected as PMS stars with strong H$\alpha$ emission in Sung et al. (2000); $^2$Object not detected in Sung et al. (2000)}}
\end{table*}

\begin{figure}
\center
\includegraphics[width=86mm, height=58mm]{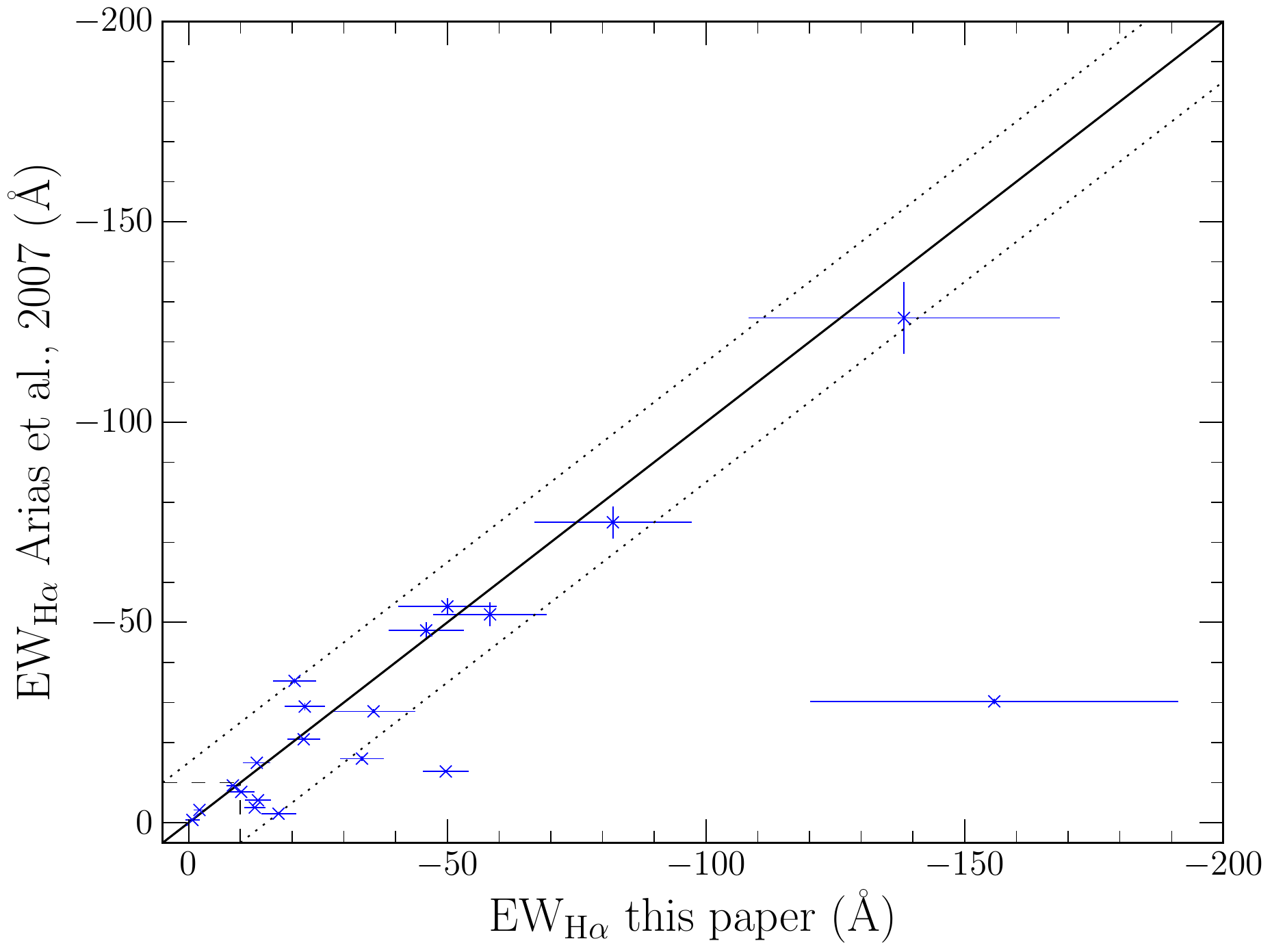}
\caption{Comparison of the photometrically determined EW$_{\rmn{H}\alpha}$ (ordinate) and spectroscopically measured EW$_{\rmn{H}\alpha}$ (abscissa) from Arias et al. (2007) for the stars selected as CTTS (crosses). The solid line represents the 1:1 relation, and the dotted lines represent the 2$\sigma$ error bars. Dashed lines represent EW$_{\rmn{H}\alpha}$\,=\,10\,\AA}
\label{fig:EWcompare}
\end{figure}


\begin{figure}
\centering
\includegraphics[width=60mm, height=60mm]{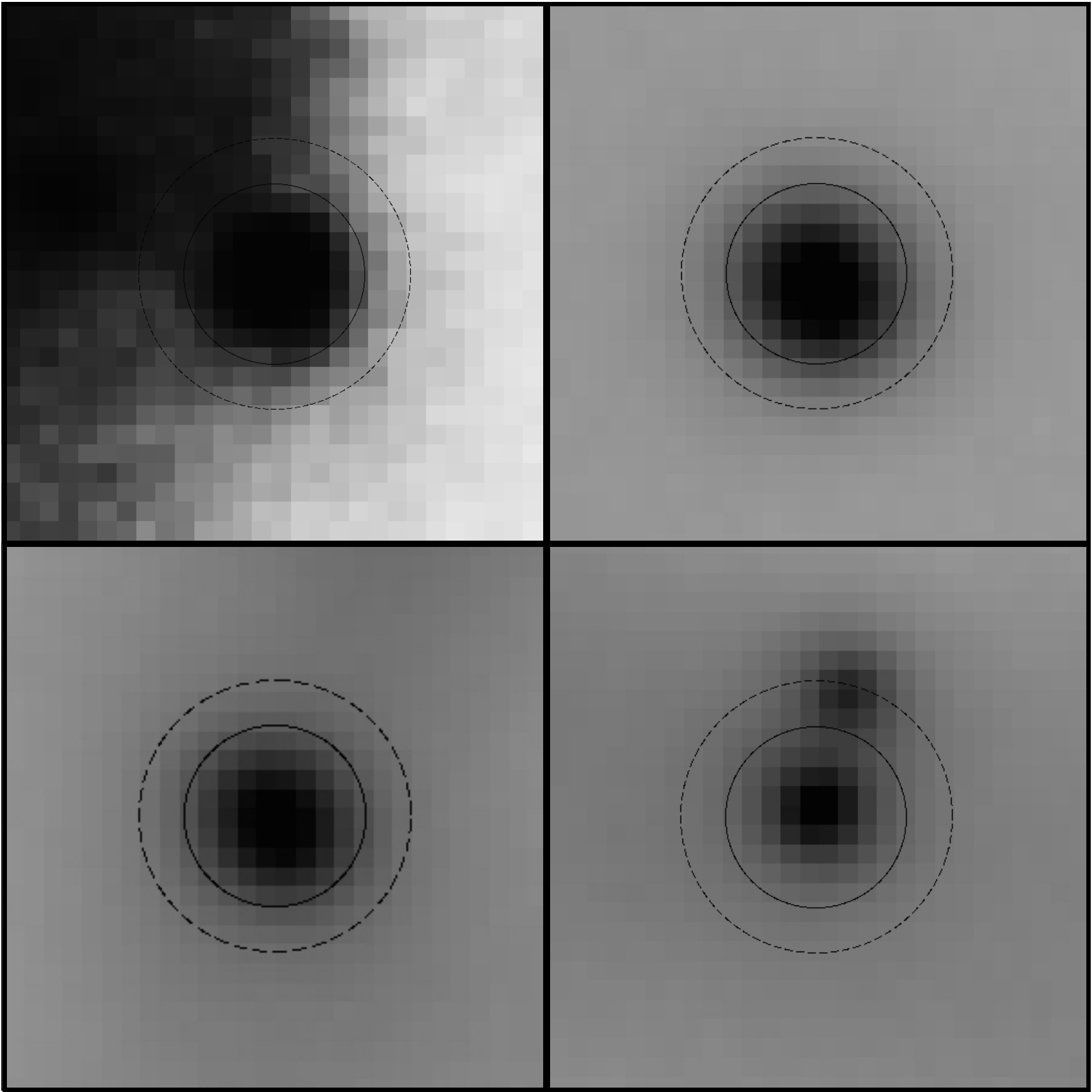}
\caption{Unsharp mask H$\alpha$ thumbnails (3$\arcsec$\,$\times$\,3\,$\arcsec$) of the three CTTS (bottom panel, upper right panel; IDs. A07\,19,30,34 in Table 3), and a H$\alpha$ emission line source that is affected by variable nebulosity (upper left panel; IDs A07\,44 in Table 3) all which have photometric EW$_{\rmn{H}\alpha}$ 2$\sigma$ greater than spectroscopic EW$_{\rmn{H}\alpha}$. The solid line marks the photometric aperture, and the dashed line the background correction aperture.}  
\label{fig:thumb}
\end{figure}

We list the spectroscopic and photometric EW$_{\rmn{H}\alpha}$ of the remaining stars in Table 3. We stress that this comparison includes all stars having $ri$H$\alpha$ photometry for completeness. Sources that are not selected as CTTS either because they do not meet the EW$_{\rmn{H}\alpha}$--spectral type selection criteria, or photometric quality checks are marked. The sample of selected CTTS does not include stars that have not met these criteria. We plot the spectroscopic and photometric EW$_{\rmn{H}\alpha}$ of the stars meeting the quality criteria in Fig.~\ref{fig:EWcompare}. We find a good correlation between the spectroscopic and photometric EW$_{\rmn{H}\alpha}$ for all objects with a median scatter $\sim$8\,\AA, approximately the error on the photometric EW$_{\rmn{H}\alpha}$ due to the assumption of uniform extinction and random photometric noise. The errors dominate the scatter for objects meeting the quality criteria and with EW$_{\rmn{H}\alpha}$~$>$~$-$10\,\AA, which is the minimum threshold used in this study, because small photometric or reddening uncertainties will produce relatively large errors in EW$_{\rmn{H}\alpha}$ for stars with weak-H$\alpha$ emission, as they lie close to the main sequence. Three CTTS show discrepant EW$_{\rmn{H}\alpha}$ ($>$\,2$\sigma$). It has been observed that the EW$_{\rmn{H}\alpha}$ of a small fraction of accretors in a given sample can vary between $-$7 to $-$100\,\AA~on time-scales greater than a year (Costigan et al. 2012) due to natural variations in the accretion process. This may be apt in the case of A07\,30/\,S00\,508 (see Table 2) as it has been marked as an eruptive variable (GCVS\,V1780\,Sgr) in \cite{kuka}. Therefore, we suggest that the difference between the spectroscopic and photometric EW$_{\rmn{H}\alpha}$ is most likely due to intrinsic accretion variability. To demonstrate that the photometric quality checks ensure that nebulous emission does not affect the measured EW$_{\rmn{H}\alpha}$ of the selected CTTS significantly, we checked the H$\alpha$ images of stars whose spectroscopic and photometric EW$_{\rmn{H}\alpha}$ differ $\sim$\,2$\sigma$, showing their object and background annuli overlaid (see Section 2.1) to ascertain the absence of rapidly spatially varying nebulosity in the background annulus (Fig.~\ref{fig:thumb}) on scales smaller than can be corrected for. For comparison a star which is affected by such nebulosity is also displayed. Finally, although steps have been taken to ensure that the results are not affected by nebulosity, we accept that due to the difficulty of separating the strong sky nebulosity from stellar emission especially in H$\alpha$ photometry, some residual may be present as is also the case in spectroscopic work in the region (Arias et al. 2007; Prisinzano et al. 2007; 2012). Due to the small number statistics of comparable spectroscopic EW$_{\rmn{H}\alpha}$, it is not possible to make a statistical statement. We therefore welcome further comparisons of the photometric EW$_{\rmn{H}\alpha}$ with spectroscopic measurements, particularly from future large surveys such as the GAIA-ESO Survey (GES; Prisinzano et al. in prep.)

\subsection{Comparison with literature H$\alpha$ emission line star lists}

We estimate the recovery rate of CTTS candidates identified using our method, and thereby the validity of our sample for a statistical analysis by comparing with all available literature optical studies of H$\alpha$ emission line stars in the region (\citealt {sung00,pris07,arias07,pris12}). \cite{sung00} observed the central NGC\,6530 region (a $\sim$20$\arcmin$ square centred on RA 18$^{h}$04$^{m}$09$^{s}$, Dec $-$24$\degr$21$\arcmin$03$\arcsec$) in {\it{UBVRI}}H$\alpha$ in a study limited to $V$\,$<$\,17\,mag. They identified 37 stars from the ($R-$H$\alpha$ )vs. ($V-I$) diagram as stars having ($R-$H$\alpha$) excess\,$>$\,0.36, which they classified as accreting PMS stars (CTTS). 
 
\begin{figure} 
\center
\includegraphics[width=86mm, height=58mm]{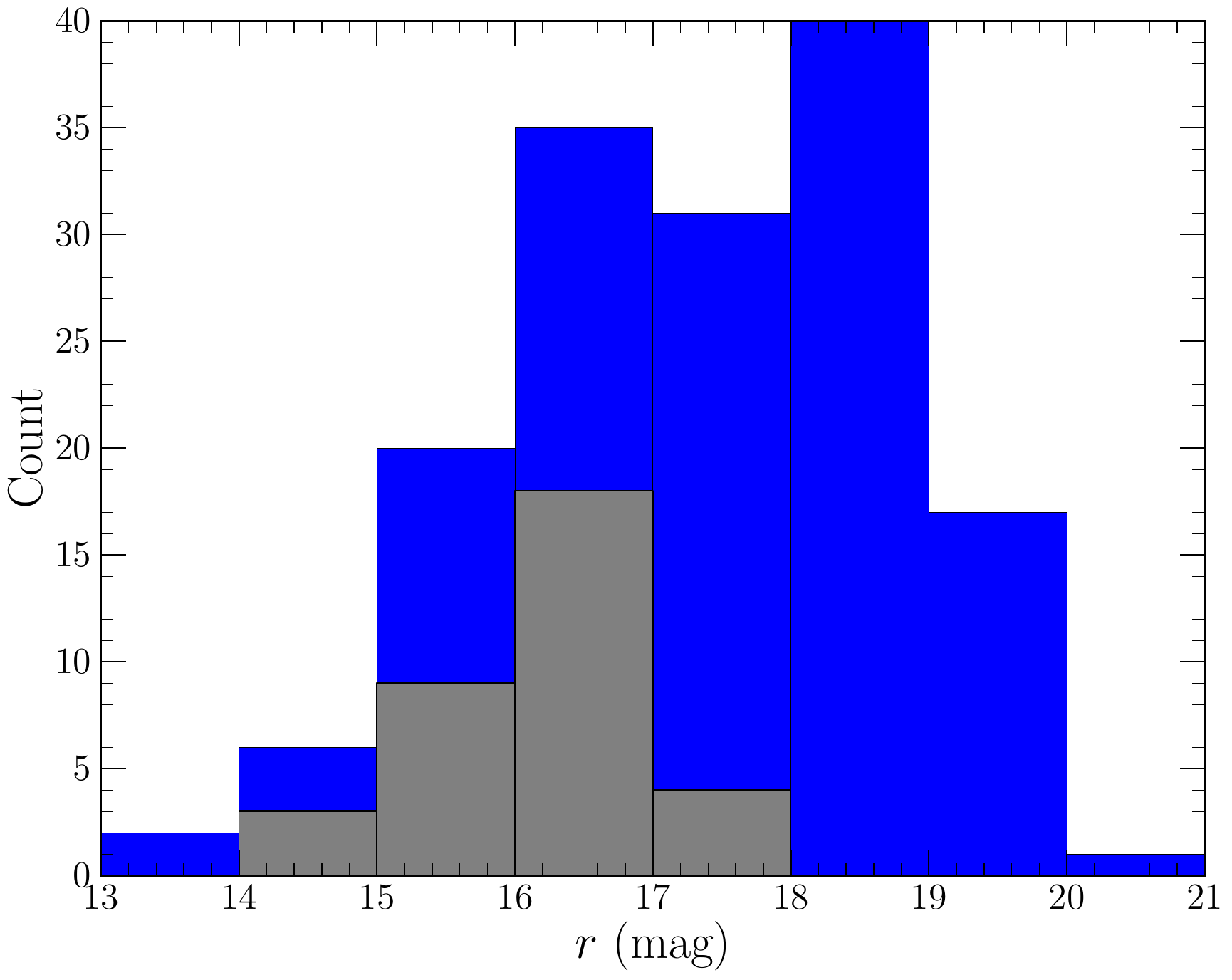}
\caption{$r$-band histogram of CTTS candidates identified in this work (blue bars) overlaid by Sung et al. (2000) CTTS candidates (grey bars). The majority of new CTTS are identified at $r$\,$<$\,17\,mag, beyond the limiting magnitude of Sung et al. (2000), while a significant fraction at magnitudes smaller than this. This is because our EW$_{\rmn{H}\alpha}$ identification method can separate CTTS lying near the main-sequence, which do not have large H$\alpha$ excesses.}
\label{fig:histcomp}
\end{figure} 

All 37 stars identified as CTTS candidates by Sung et al. (2000) have VPHAS+ photometry, although four sources are saturated and their photometric EW$_{\rmn{H}\alpha}$ may be inaccurate. The remaining stars have photometric EW$_{\rmn{H}\alpha}$\,$<$\,$-$8\,\AA~indicating that they are reliably identified as emission-line stars using VPHAS+ photometry. Additionally, in the region of the sky observed by \cite{sung00} we identify 125 new CTTS candidates. This is roughly three times as many as identified by Sung et al. (2000). A comparison of the magnitude distribution of the two samples is plotted in Fig.~\ref{fig:histcomp}. We converted Sung et al. (2000) $RI$ magnitudes to VPHAS+ $r$-band using the colour equations of \cite{jester05}. A significant fraction of VPHAS+ identified CTTS candidates are located at $r$\,$>$\,17\,mag, beyond the limiting magnitude of \cite{sung00}. We roughly double the number of Sung et al. (2000) CTTS candidates at $r$\,$<$\,16\,mag, as we can separate them from main sequence stars on the basis of their EW$_{\rmn{H}\alpha}$, without needing to rely on large ($R-$H$\alpha$) excesses.

\begin{figure} 
\center
\includegraphics[width=86mm, height=58mm]{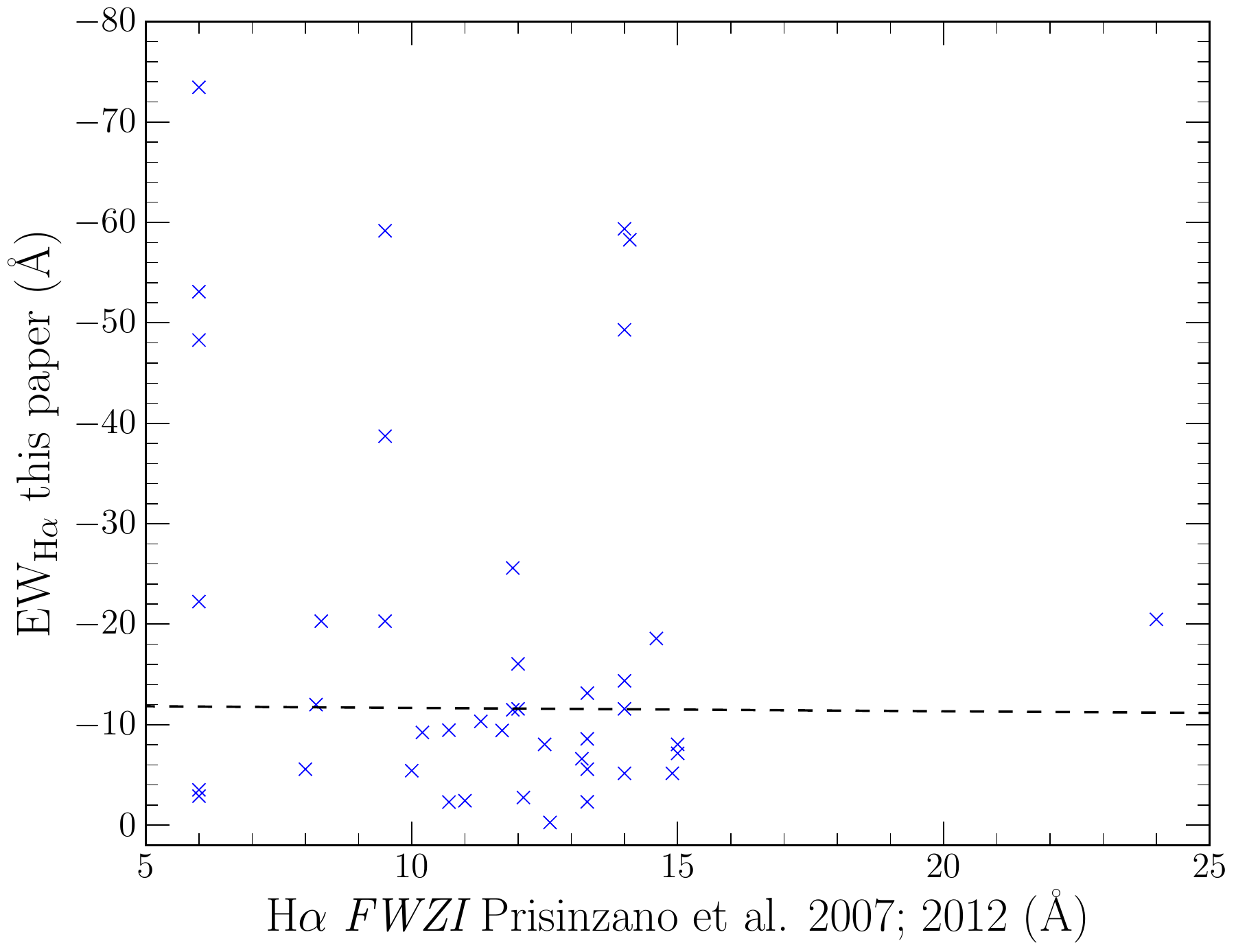}
\caption{{\it{{\it{FWZI}}}} of H$\alpha$ line profiles measured by Prisinzano et al. (2007; 2012) plotted against EW$_{\rmn{H}\alpha}$ measured using the photometric method. All of the stars plotted were classified as CTTS by Prisinzano et al. (2007;2012) based on the measured H$\alpha$ {\it{{\it{FWZI}}}} and line profile. We find that almost all these stars are in emission in the VPHAS+ sample, but a significant majority of them are below the minimum EW$_{\rmn{H}\alpha}$ threshold for CTTS (dashed line). We find no relation between H$\alpha$ {\it{{\it{FWZI}}}} and EW$_{\rmn{H}\alpha}$.}
\label{fig:comppris}
\end{figure} 

\cite{pris07} presented H$\alpha$ fibre spectra of 95 stars in the Lagoon Nebula, but experienced difficulty in separating the stellar H$\alpha$ component from any nebular contamination. They do not measure stellar EW$_{\rmn{H}\alpha}$ to classify accreting PMS stars, instead employing the Full Width Zero Intensity ({\it{{\it{FWZI}}}}) of the H$\alpha$ line to do so. They argue that broad emission due to accretion can be separated from any narrow nebular contamination using the {\it{FWZI}} parameter. \cite{pris07} classify 31 of the 95 stars for which they present H$\alpha$ spectra as CTTS (Fig.~\ref{fig:comppris}) {\it{purely}} based on the {\it{FWZI}} value if the H$\alpha$ line is in emission. 30 are identified as having H$\alpha$ in emission in our sample, although slightly less than half are below the spectral type-EW$_{\rmn{H}\alpha}$ selection criteria we apply to identify CTTS. We find no correlation between the H$\alpha$ {\it{FWZI}} and EW$_{\rmn{H}\alpha}$. \cite{pris12} used new H$\alpha$ multi-slit observations to add a further 16 accreting stars to the sample of \cite{pris07}, which are also shown in Fig.~\ref{fig:comppris}. All of these stars have H$\alpha$ in emission, but around a third of the sample fall below the spectral type-EW$_{\rmn{H}\alpha}$ selection criteria. In short, our method can reliably identify H$\alpha$ emission in all stars that have been previously classified as H$\alpha$ emission-line stars using spectra. We stress that although the {\it{FWZI}} of the H$\alpha$ line is a good identifier of accretion, but may not be an accurate measure of the accretion rate in CTTS (\citealt{white03, cost14}).

The comparison of our sample with literature H$\alpha$ emission line star lists suggest that all known CTTS sources within the VPHAS+ dynamic range can be reliably recovered. We identify a total of 235 CTTS candidates over our area of study, which contains at least 170 new accreting sources that represent $\sim$ 200\,per\,cent increase in known CTTS stars within the region. The area covered by this study is significantly larger (around 3.5 times) than previous photometric searches for PMS stars in the Lagoon Nebula (see Tothill et al. 2008 for a summary). We conclude that the uniform expansive sky coverage, and photometric depth afforded by the VPHAS+ survey can homogeneously identify a substantial number of CTTS candidates; and the photometric method estimates the EW$_{\rmn{H}\alpha}$ accurately in comparison with intermediate-resolution spectroscopy.

\section{Results}

In the current accretion paradigm, it is assumed that the stellar magnetosphere truncates the circumstellar disc at an inner radius (\citealt{hart94}; \citealt*{muzz98}). Gas flows along the magnetic field lines at this truncation radius, releasing energy when impacting on the star. It is generally accepted that the accretion energy goes towards heating and ionising the gas, and that the accretion luminosity ($L_{\rmn{acc}}$) can be measured from the reradiated emission line luminosity (\citealt{calv98, hart08}). The $\dot{M}_{\rmn{acc}}$ can then be estimated from the free-fall equation: 
  \begin{equation}
    \dot{M}_{\rmn{acc}}= \frac{L_{\rmn{acc}}R_{\ast}}{\rmn{G}M_{\ast}}\Bigg(\frac{R_{\rmn{in}}}{R_{\rmn{in}}-R_{\ast}}\Bigg).
  \end{equation} 
$M_{\ast}$ and $R_{\ast}$ are the stellar mass and radius respectively. $R_{\rmn{in}}$ is the truncation radius. 

In this Section, we estimate the stellar and accretion properties of our candidate CTTS based on the magnetosphere accretion paradigm. In Section 4.1 we estimate their stellar masses and ages. We estimate their $\dot{M}_{\rmn{acc}}$ based on their EW$_{\rmn{H}\alpha}$ in Section 4.2 and $u$-band excess in Section 4.3. We discuss their near-infrared properties in Section 4.4. The caveats, completeness and contamination of our sample are discussed in Sections 4.5-4.8.

\subsection{Stellar properties}

\subsubsection{Mass and age}
\begin{figure*}
\center
\includegraphics[width=150mm, height=125mm]{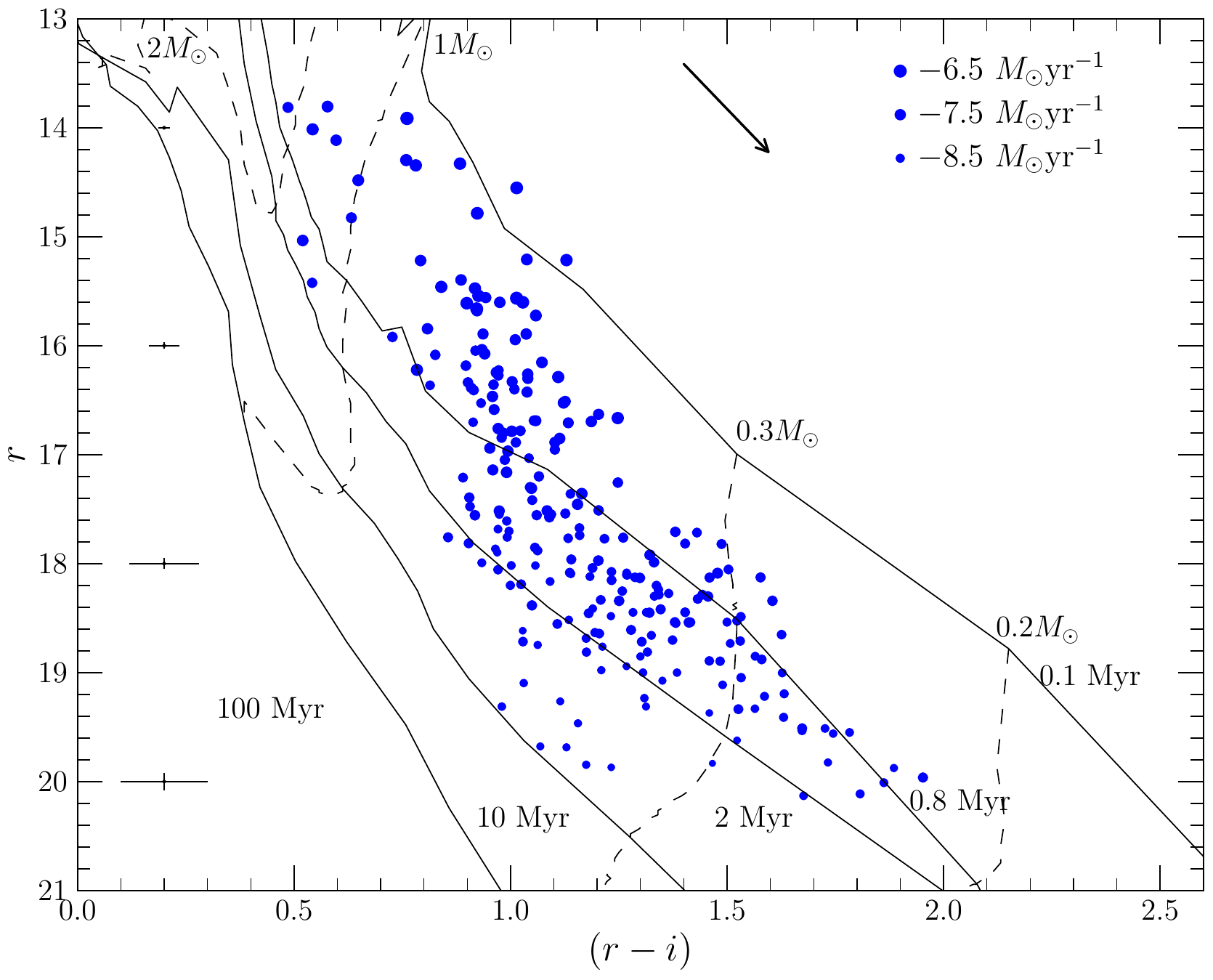}
\caption{$r$ versus ($r$-$i$) CMD of CTTS candidates. Labelled isochrones (solid lines) and tracks (dashed lines) are from Siess et al. (2000). Model isochrones and tracks are reddened by $E(B-V)$\,=\,0.35 following a reddening law $R_{V}$\,=\,3.1. Circles indicate the position of each candidate CTTS, and symbol size is representative of log\,$\dot{M}_{\rmn{acc}}$ as shown by the legend in the top-right hand corner. The reddening vector for $A_{V}$\,=\,1 is shown. The median magnitude errors in a given 2\,mag bin, and the associated colour errors are shown at the corresponding magnitude and ($r-i$)\,=0.02.}
\label{fig:cmd}
\end{figure*}

The mass ($M_{*}$) and age ($t_{*}$) of each star is estimated by interpolating its position in the observed $r$ versus $(r-i)$ CMD with respect to PMS tracks and isochrones (Fig.~\ref{fig:cmd}). \cite*{siess00} solar metallicity single star isochrones and tracks were employed as they cover the mass and age range of our sample. The isochrones were reddened assuming the mean reddening $E(B-V)$\,=\,0.35 following a reddening law $R_{V}$\,=\,3.1 (see Section 1). The systematic differences that would arise on adopting different stellar models are discussed in Section 4.6.1. The tracks and isochrones were converted from the luminosity- temperature plane to the colour-magnitude plane using the colour equations of \cite{jester05}.
 
Fig.~\ref{fig:hist}(a) shows the luminosity function of our sample. The interpolated mass distribution of our candidate CTTS is plotted in Fig.~\ref{fig:hist}(b). All our candidate CTTS have masses between 0.2\,-\,2.2\,$M_{\odot}$. The mass distribution shows the expected increase towards lower masses, peaking at 0.35\,$M_{\*}$. The mass cut-off and steady decline in sub-solar mass below 0.3\,$M_{\odot}$ is expected because of the limiting magnitude. 

The ages of stars in our sample range between 0\,-\,9\,Myr, with 95\,per\,cent of stars having an age less than 2\,Myr (Fig.~\ref{fig:hist} b). Few stars fall beyond the 30000\,yr isochrone. For these stars, age estimates represent an upper limit. The median age of our sample is 0.85\,$\pm$\,0.6\,Myr. This agrees with the main-sequence lifetime of the early-type stars found in the region \citep{tothill08}. In particular, 9Sgr (O4\,V; \citealt{wal73}) and H36 (O7V; van Altena \& Jones 1972) are two prominent O-type members still lying on the zero-age main sequence (ZAMS), and are expected to evolve off the ZAMS after 1-3\,Myr. Chen et al. (2007) measured an intrinsic velocity dispersion of 8\,km\,s$^{-1}$. Assuming this as an expansion velocity, Tothill (2008) calculated a dynamical age of 1\,Myr. Furthermore, the age of the PMS stars that have X-ray emission (Damiani et al. 2004) or H$\alpha$ emission (Sung et al. 2000; Arias et al. 2007) are estimated to be around 1\,Myr.

We caution the reader that higher mean age estimates ($>$\,2\,Myr) for the general population of stars in the Lagoon Nebula have been measured in the literature. In particular, Walker (1957) and van Alten \& Jones (1972) suggest an age estimate $\sim$\,3\,Myr, although both of these studies only fit a ZAMS and use a distance modulus of around 11.5\,mag. Recent works by Mayne et al. (2007) and Mayne \& Naylor (2008) use the CMD diagram to build a relative age ladder, and obtain an age of 2\,Myr. Also, van den Ancker et al. (1997) find a highly probable member with an age of 15\,Myr and suggest a prolonged period of star formation taking place within the Lagoon Nebula. The evidence for this remains unclear. The study by Prisinzano et al. (2005) discussed the possibility of an age spread, by comparing their optical sample with the X-ray identified PMS stars of Damiani et al. (2004). They find a median age of 2\,Myr using their optical sample, with the older stars preferentially located to the northeast, and younger stars located to the south and west. In our sample, higher or lower ages may be estimated for a few stars because the assumption of uniform distance or extinction may be inappropriate, or because they are non-members and not representative of the age of the sample. Since our study is limited to an assumption of uniform extinction and no strict membership tests barring the EW$_{\rmn{H}\alpha}$ criteria, we cannot differentiate between ages that are incorrect due to these effects or genuinely older. Therefore, we cannot confirm or rule out the presence of a true age spread, and further data on cluster membership is required to explore this possibility. 

Finally, we stress that the precise ages of individual stars derived using isochrones may be uncertain, as most of our stars are around one million years old. At this point in a star's evolution, age determinations using stellar isochrones are imprecise because the age is thought to be still dependant on the accretion history and stellar birthline corrections. However, the median age of the CTTS population ($\sim$\,1\,Myr) is relatively accurate given large number statistics, and the fact our result agrees well with age estimations based on the expansion velocity (Chen et al. 2007) main-sequence lifetime of OB stars, and literature age estimations of the pre-main sequence population (\citealt{sung00, dami04, arias07}).

\subsection{Accretion properties}

\subsubsection{H$\alpha$ line luminosity}
The H$\alpha$ line flux ($F_{\rmn{H}\alpha}$) is calculated by subtracting the stellar continuum flux ($F_{\rmn{continuum}}$) from the total flux in the H$\alpha$ band ($F_{\rmn{total}}$). The total H$\alpha$ flux ($F_{\rmn{total}}$) is given by the unreddened H$\alpha$ magnitude ($m_{\rmn{H}\alpha}$) according to the relation,
  \begin{equation}
    \ F_{\rmn{total}}= F_{0}\times[10^{-0.4(m_{\rmn{H}\alpha}+0.03)}].\
  \end{equation}
$m_{\rmn{H}\alpha}$ is the unreddened H$\alpha$ magnitude. $F_{0}$ is the band-integrated reference flux. For the VPHAS+ $\rmn{H}\alpha$ filter, $F_{0}$=1.84$\times10^{-7}$~ergs\,$\rmn{cm}^{2}$\,$\rmn{s}^{-1}$. 
$F_{\rmn{total}}$ is a sum of the continuum ($F_{\rmn{continuum}}$) and line flux ($F_{\rmn{H}\alpha}$). 
  \begin{equation}
F_{\rmn{H}\alpha} =F_{\rmn{total}} - F_{\rmn{continuum}}. 
  \end{equation}
Following from equation 1, equation 4 can be written as
  \begin{equation} 
F_{\rmn{H}\alpha} = F_{\rmn{total}}\times\frac{-\rmn{EW}_{\rmn{H}\alpha}/{\rmn{W}}}{1-\rmn{EW}_{\rmn{H}\alpha}/{\rmn{W}}}.
  \end{equation}
The line luminosity is given by $L_{\rmn{H}\alpha}$:
  \begin{equation} 
L_{\rmn{H}\alpha} = F_{\rmn{H}\alpha}\times4\pi d^{2}.
  \end{equation}

Here the distance $d$\,=\,1250\,$\pm$\,50\,pc. Contamination from N[{\scriptsize II}] $\lambda$$\lambda$ 6548, 6584 is excluded by assuming it causes 2.4 per\,cent of the H$\alpha$ intensity \citep{de10}.

\subsubsection{Accretion luminosity ($L_{\rmn{acc}}$)}

Measurements of $L_{\rmn{acc}}$ from UV continuum spectral modelling in the literature (\citealt*{sicilia10}; \citealt{herc08, dahm08, hart03}) were plotted against $L_{\rmn{H}\alpha}$ by \cite{geert11}. A power law relationship, $L_{\rmn{acc}}$ $\propto$ ${L_{\rmn{H}\alpha}}^{\gamma}$ was found, having an index $\gamma$\,=\,1.13, consistent with the accretion models of \cite{muzz98}. Based on results of Barentsen et al. (2011), we assume:
\begin{equation} 
\rmn{log}~L_{\rmn{acc}} = (1.13\pm 0.07)\rmn{log}~L_{\rmn{H}\alpha}+(1.93\pm 0.23)
\end{equation}
$\rmn{log}~L_{\rmn{acc}}$ and $\rmn{log}~L_{\rmn{H}\alpha}$ are measured in units of solar luminosity, $L_{\odot}$. The root mean scatter~=~0.54. The scatter in the observed relationship is likely caused by a combination of circumstellar absorption, uncertain extinction and line emission caused by processes other than accretion.

\subsubsection{Mass accretion rate $\dot{M}_{\rmn{acc}}$}

\begin{figure*}
\center
\includegraphics[width=100mm, height=80mm]{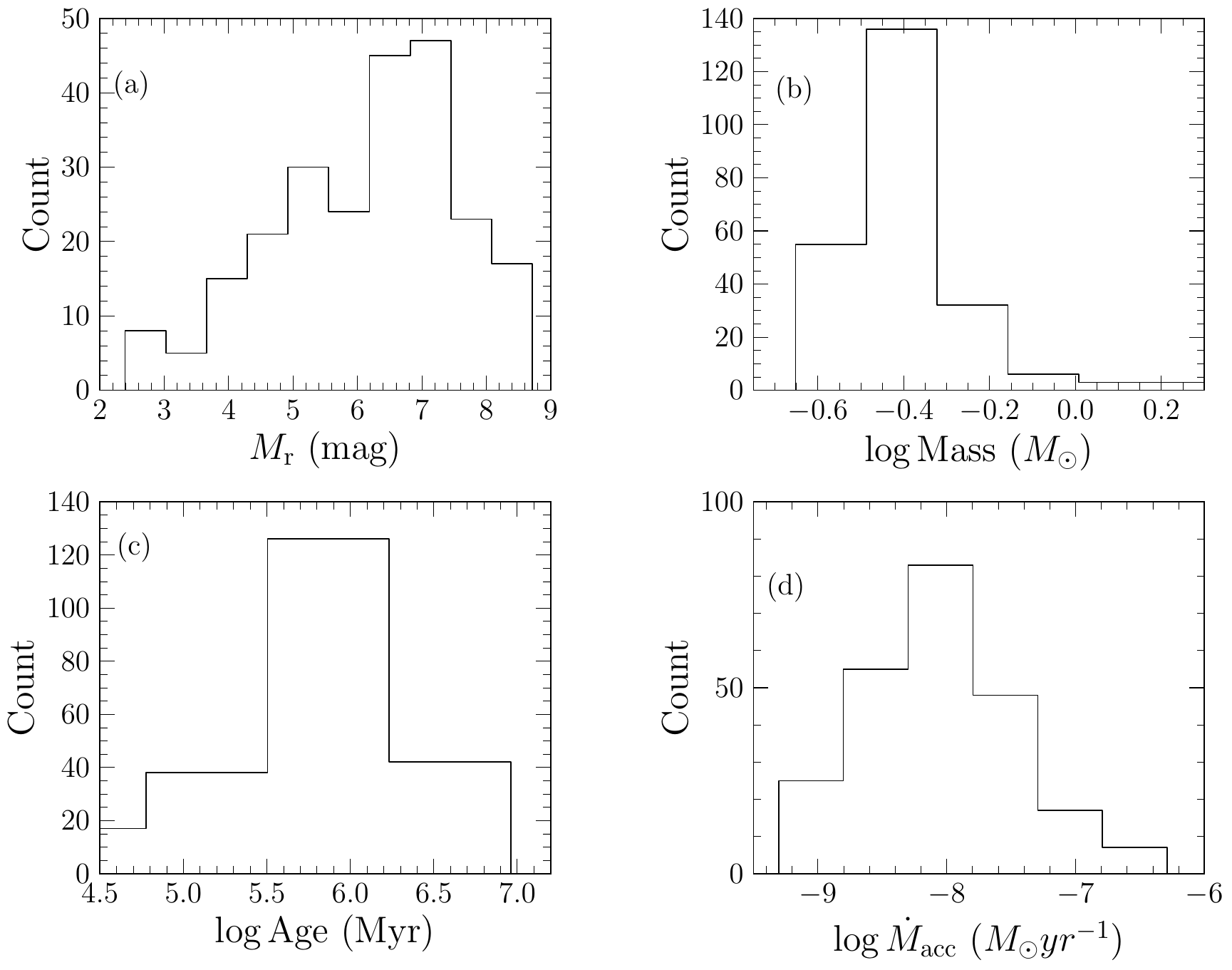}
\caption{Distribution of magnitude $M_{\rmn{r}}$ (a), Mass (b), Age (c) and $\dot{M}_{\rmn{acc}}$ (d) of our sample.}
\label{fig:hist}
\end{figure*}

$\dot{M}_{\rmn{acc}}$ are estimated from Equation 2. We adopt $R_{\rmn{in}}$\,=\,5\,\,$\pm$\,\,2\,$R_{\ast}$ 
from (\citealt{gullbring98,vink05}). The resultant distribution of mass accretion rates are plotted in Fig.~\ref{fig:hist}(d). The median $\dot{M}_{\rmn{acc}}$ is 10$^{-8.1}$\,$M_{\odot}yr^{-1}$. Our results are given in Table 4.

\subsection{Comparison between $u$-band and H$\alpha$ mass accretion rates}

\begin{figure*}
\center
\includegraphics[width=175mm, height=65mm]{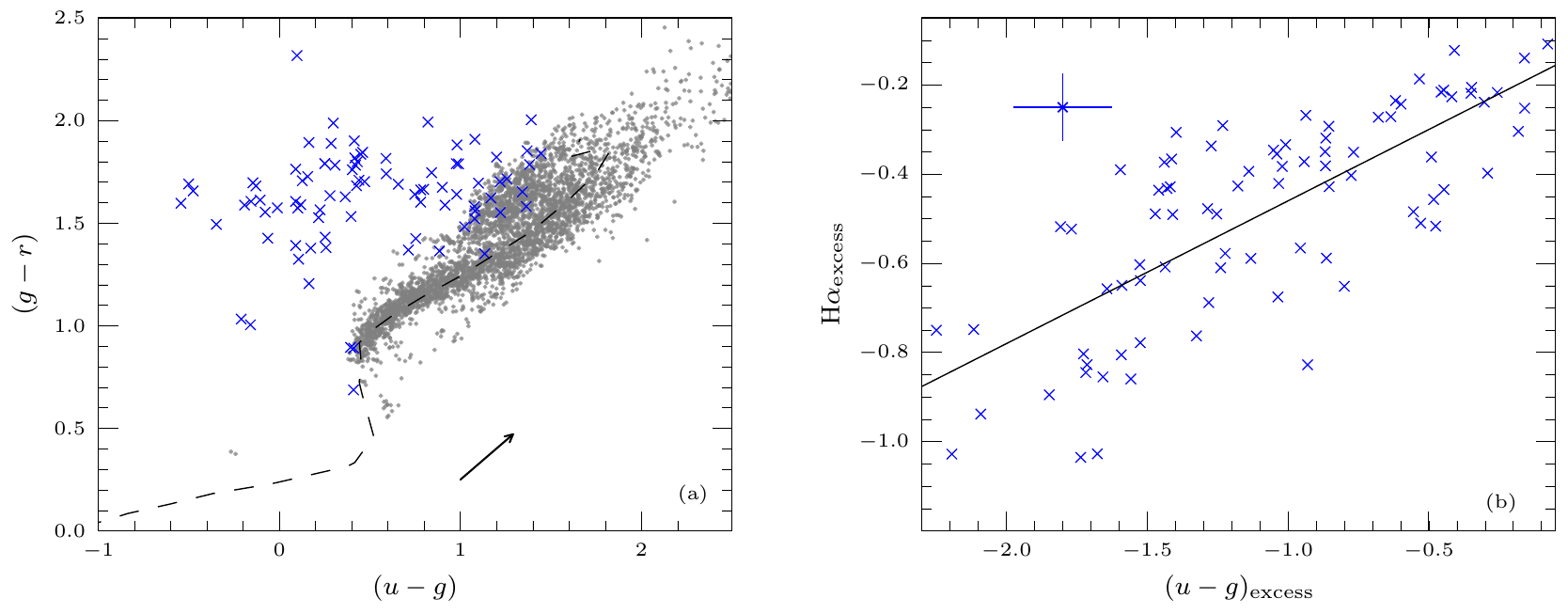}
\caption{(a) ($u-g$) versus ($g-r$) diagram. The dashed line is the reddened model colour track from Castelli \& Kurucz (2004). Stars meeting our selection criteria (grey dots) and CTTS identified on the basis of their EW$_{\rmn{H}\alpha}$ (blue crosses) are shown. A reddening vector for $A_{V}$\,=\,1 is shown. (b) Comparison of $(u-g)_{\rmn{excess}}$ and H$\alpha_{\rmn{excess}}$ of stars in (a). The solid line represent the best-fit relation having H$\alpha_{\rmn{excess}}$\,=\,0.3\,$(u-g)_{\rmn{excess}}$\,$-$\,0.14.}
\label{fig:ugr} 
\end{figure*}

91 candidate CTTS have $ugr$ photometry from the blue data set meeting our selection criteria. Four stars do not show any measurable $UV$-excess in the ($u-g$) vs. ($g-r$) diagram, and have EW$_{\rmn{H}\alpha}$\,$>$\,$-$20\,\AA. We discuss them in more detail in Section 4.8. For the remaining stars, we measure $\dot{M}_{\rmn{acc}}$ from the excess ultraviolet emission \citep{gullbring98}.  

\begin{figure}
\center
\includegraphics[width=86mm, height=58mm]{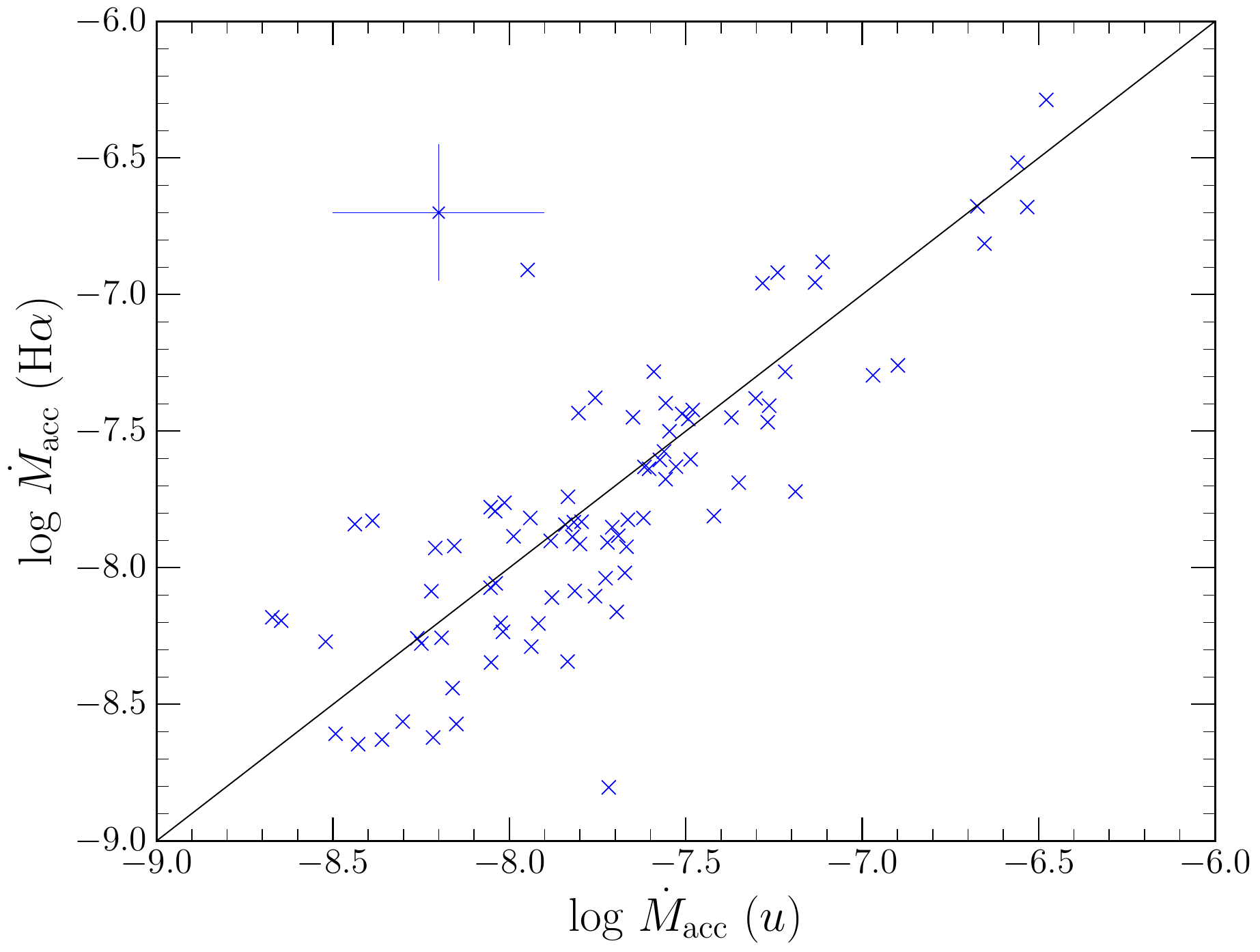}
\caption{Comparison of $\dot{M}_{\rmn{acc}}$ derived from H$\alpha$ line luminosity and $u$-band excess luminosity. Mean error bars are shown in the upper left hand corner.}
\label{fig:ugr1} 
\end{figure}

Similar to the method described in Section 3, the $(g-r)$ colour is used to estimate spectral type, while the $(u-g)$ colour is the measure of UV-excess due to the accretion process. The UV-excess stars are shown in Fig.~\ref{fig:ugr}(a). The excess flux in the $u$-band ($F_{u,\rmn{excess}}$) is: 

  \begin{equation}
    \ F_{u,\rmn{excess}}= F_{0,u}\times[10^{-u_{0}/2.5}-10^{-(u-g)_{\rmn{model}}+g_{0})/2.5}].\
  \end{equation}
$F_{0,u}$ is the $u$-band integrated reference flux. $u_{0}$ and $g_{0}$ are the observed dereddened magnitudes, and $(u-g)_{model}$ is the corresponding model colour from \cite{cast04}. The excess flux is used to estimate the $L_{\rmn{acc}}$ using the empirical relation \citep{gullbring98};
\begin{equation} 
\rmn{log}~L_{\rmn{acc}} = {log}~L_{u,\rmn{excess}}+0.98.
\end{equation}
$L_{u,\rmn{excess}}$ is the $u$-band excess luminosity. $L_{\rmn{acc}}$ and $L_{u,\rmn{excess}}$ are in units of solar luminosity $L_{\odot}$. Mass accretion rates are calculated using Eq.~2, with results given alongside H$\alpha$ derived $\dot{M}_{\rmn{acc}}$ in Table 4. The $(u-g)_{\rmn{excess}}$ are compared with the H$\alpha_{\rmn{excess}}$ in Fig.~\ref{fig:ugr}(b), and $\dot{M}_{\rmn{acc}}$ thus determined are compared with the ones determined from the H$\alpha$-line luminosity in Fig.~\ref{fig:ugr1}.

The $(u-g)_{\rmn{excess}}$ and H$\alpha_{\rmn{excess}}$, two independent (except for the assumed extinction value) measures of accretion correlate well. We find the best fit relation to follow H$\alpha_{\rmn{excess}}$\,=\,0.3\,$(u-g)_{\rmn{excess}}$\,$-$\,0.14 (Fig.~\ref{fig:ugr}b). There are only six low accretion rate objects (log\,$\dot{M}_{\rmn{acc}}$\,$<$\,$-$8.5) detected in the {\it u}-band, that are detected in H$\alpha$ (Fig.~\ref{fig:hist}d). This is due to the faint magnitude limit in $u$ $\sim$ 19\,mag, thereby excluding the low-mass stars detected in H$\alpha$. The logarithm of the mass accretion rates determined from the H$\alpha$-line luminosity and $u$-band excess luminosity have a mean variation (R.M.S. scatter) of 0.17\,dex. There exist a minority of CTTS candidates (four) which have differences in log\,$\dot{M}_{\rmn{acc}}$ $\sim$\,0.8\,dex. Since the observations are non-contemporaneous, the $\dot{M}_{\rmn{acc}}$ should display intrinsic variabilities over long time-scales with maximum variations in log\,$\dot{M}_{\rmn{acc}}$ expected to be $\lesssim$\,0.52\,dex \citep{cost12}.

This result indicates no significant statistical effect of accretion variability on the sample as a whole. However, accurate $u$ band photometry of low-mass stars is comparatively more challenging to obtain. Therefore H$\alpha$ photometry is better suited to provide statistically significant samples of $\dot{M}_{\rmn{acc}}$.

\begin{table*}
\caption{Derived properties of candidate CTTS in the Lagoon Nebula. The full table is available in the online version of this journal.}
\centering
\begin{tabular}{|l|l|l|l|l|l|}
\hline
  \multicolumn{1}{c}{JNAME} &
  \multicolumn{1}{c}{EW$_{{\text{H}}\alpha}$} &
  \multicolumn{1}{c}{Mass} &
  \multicolumn{1}{c}{Log\,Age} &
  \multicolumn{1}{c}{$\dot{M}_{\rm{acc}}$$_{,{\text{H}}\alpha}$} &
  \multicolumn{1}{c}{$\dot{M}_{\rm{acc,\, U}}$} \\
  \multicolumn{1}{c}{} &
  \multicolumn{1}{c}{(\AA)} &
  \multicolumn{1}{c}{($M_{\odot}$)} &
  \multicolumn{1}{c}{{yr}} &
  \multicolumn{1}{c}{($M_{\odot}yr^{-1}$)} &
  \multicolumn{1}{c}{($M_{\odot}yr^{-1}$)} \\
\hline
\hline
  18023206$-$2415320 & $-$14.34$\pm$2.66 & 1.0$\pm$0.00 & 5.6$\pm$0.08 & $-$7.7$\pm$0.6 & \\
  18023355$-$2414548 & $-$18.90$\pm$3.27 & 0.4$\pm$0.01 & 4.7$\pm$0.19 & $-$7.9$\pm$0.6 & \\
  18023368$-$2418022 & $-$213.96$\pm$36.9 & 0.2$\pm$0.20 & 5.5$\pm$2.54 & $-$8.0$\pm$0.6 & \\
  18023560$-$2413024 & $-$63.16$\pm$10.8 & 0.3$\pm$0.05 & 5.4$\pm$0.69 & $-$8.1$\pm$0.6 & $-$7.8$\pm$0.7\\
  18023730$-$2416242 & $-$28.39$\pm$5.15 & 0.4$\pm$0.01 & 4.6$\pm$0.20 & $-$7.6$\pm$0.6 & \\
  18023836$-$2419313 & $-$36.71$\pm$4.95 & 0.4$\pm$0.06 & 6.2$\pm$0.85 & $-$8.8$\pm$0.6 & $-$7.7$\pm$0.7\\
  18023897$-$2414274 & $-$110.99$\pm$21.0 & 0.5$\pm$0.04 & 6.3$\pm$0.54 & $-$8.1$\pm$0.6 & \\
  18023972$-$2419310 & $-$33.25$\pm$4.58 & 0.3$\pm$0.06 & 5.6$\pm$0.77 & $-$8.6$\pm$0.6 & $-$8.2$\pm$0.7\\
  18024000$-$2419346 & $-$31.36$\pm$5.14 & 0.4$\pm$0.03 & 4.7$\pm$0.43 & $-$8.3$\pm$0.6 & $-$7.9$\pm$0.7\\
  18024097$-$2412163 & $-$39.45$\pm$7.23 & 0.5$\pm$0.01 & 5.7$\pm$0.23 & $-$7.7$\pm$0.6 & $-$7.2$\pm$0.7\\
\hline\\
\end{tabular}
\end{table*}

\subsection{Infrared properties}
A simple sanity check on the validity of the identified CTTS can be made by studying their near-infrared properties. This is because CTTS display near-infrared excesses when compared to a purely stellar template due to the presence of dust in their inner circumstellar discs \citep{cohen79}.

Near-infrared {\it{JHK}} photometry in the Lagoon Nebula from the UKIRT Infrared Deep Sky Survey (UKIDSS) Galactic Plane survey (\citealt{ukidss}) was cross-matched with the positions of the CTTS in our sample. Only cross-matches having random error $<$\,0.1\,mag in {\it{JHK}}, and classified as having a stellar profile in the UKIDSS catalogue were selected. UKIDSS reaches a depth of 5$\sigma$ at $K$\,$\sim$\,19\,mag, which translates to roughly a star of spectral type M6 in the Lagoon Nebula, suggesting that most of the stars in our sample should be identifiable in the UKIDSS survey. 224 (95\,per\,cent) stars in our sample were found to have UKIDSS cross-matches. 

\begin{figure*}
\center
\includegraphics[width=175mm, height=65mm]{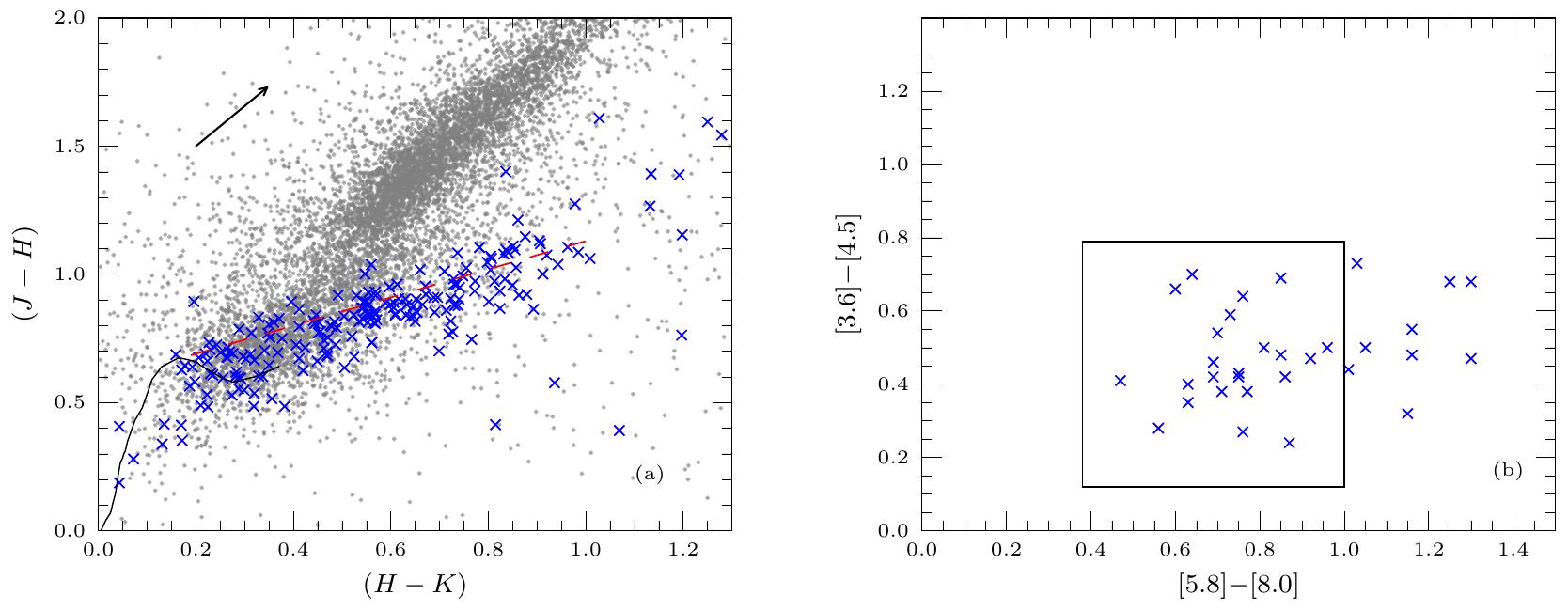}
\caption{(a) ($J-H$) vs. ($H-K$) colour-colour diagram. Stars located in the line of sight of the Lagoon Nebula having UKIDSS photometry (grey dots), and cross-matched CTTS candidates (blue crosses) are shown. The red dashed line represents the CTTS locus of Meyer et al. (1997). The solid line is the main sequence locus from Bessell et al. (1998). The reddening vector for $A_{V}$\,=\,1 is shown. (b) The CTTS candidates having cross matches in {\it{Spitzer}} photometry are shown as blue crosses. The box outlines the predicted colours for PMS stars having $\dot{M}_{\rmn{acc}}$\,$\sim$\,$10^{-8}$\,$M_{\odot}yr^{-1}$ (D'Alessio et al. 2005).}
\label{fig:ir}
\end{figure*}

In Fig.~\ref{fig:ir}(a), the ($J-H$) vs. ($H-K$) colour-colour diagram of stars located in the Lagoon nebula is plotted. Also shown are near-infrared counterparts of cross-matched candidate CTTS. The dashed line is the CTTS locus of \cite*{meyer97} which predicts the excess emission using disc accretion models having log\,$\dot{M}_{\rmn{acc}}$ between $-$6 to $-8$\,$M_{\odot}yr^{-1}$. The solid line is the main sequence locus from \cite*{bess98}. Most stars ($\sim$\,85\,per\,cent) having near-infrared colours in our sample lie on the CTTS locus, suggesting on the basis of the near-infrared diagram alone that they are CTTS. This confirms our sample as consisting primarily of CTTS undergoing accretion. Around 35 stars lie near the main sequence locus, indicating that they are either weak-line T\,Tauri stars, or have no excess infrared emission suggestive of a circumstellar disc. Three of the stars lying below the main sequence spur are the stars that showed no $UV$-excess. We discuss them further in Section 4.8.  

We also cross-matched our sample with mid-infrared {\it{Spitzer}} data \citep{dewa10}. 32 cross-matches were found. In Fig.~\ref{fig:ir}(b), the 3.6\,$\micron-$4.5\,$\micron$ vs. 5.8\,$\micron-$8\,$\micron$ colour-colour diagram is plotted. The solid box represents the typical colours for PMS stars having mean $\dot{M}_{\rmn{acc}}$\,=\,$10^{-8}$\,$M_{\odot}yr^{-1}$ \citep{d05}. Most of the cross-matched CTTS candidates in our sample lie in this box, providing an additional sanity check on the results. Six stars lie at a 5.8\,$\micron$-8\,$\micron$ colour much brighter than maximum expected colour. Their accretion rates are approximately a magnitude higher than the model at $\sim$\,$10^{-7}$\,$M_{\odot}yr^{-1}$.

\subsection{Random errors}

The random errors affecting the determined parameters are (i) random photometric noise (ii) scatter in the extinctions of individual stars compared to the adopted uniform value
(iii) scatter in the distances of individual stars compared to the adopted uniform value (iv) scatter in the $L_{\rmn{H}\alpha}$-$L_{\rmn{acc}}$ relation (Eq. 7).

The errors due to random photometric noise are propagated through the results. The true extinction for some stars may vary from the assumed uniform extinction. To calculate the errors arising due to the assumption of uniform extinction, we assume $\sigma_{E(B-V)}$\,=\,0.2 from Section 1. The true distance for some stars may vary from the assumed distance. To account for the error due to the uncertainty in the distance, we assume a $\sigma_{d}$\,=\,50\,pc. 50\,pc is the width of the region at the assumed distance. The RMS scatter of 0.54 in the $L_{\rmn{H}\alpha}$-$L_{\rmn{acc}}$ relation was used to calculate errors arising due to the deviation from this relationship for individual objects.

The mean errors ($\Bar{\sigma}$) determined are listed in Table~5. We find that our EW$_{\rmn{H}\alpha}$ is accurate within 7\,\AA~while taking into account random photometric and extinction uncertainties. The stellar mass and age are constrained to 0.08\,$M_{\odot}$ and 0.6\,Myr respectively accounting for random photometric, extinction and distance uncertainties. The error on $\dot{M}_{\rmn{acc}}$ is dominated by the scatter in the $L_{\rmn{H}\alpha}$-$L_{\rmn{acc}}$ relation.   

\begin{table}
\caption{Errors in the derived parameters.}
\label{tab:error}
\centering{%
\begin{tabular}{|l|l|l|l|}
\hline
  \multicolumn{1}{|c|}{Parameter} &
  \multicolumn{1}{c|}{Median} &
  \multicolumn{1}{c|}{Range} &
  \multicolumn{1}{c|}{$\Bar{\sigma}$} \\
\hline
  EW$_{\rmn{H}\alpha}$ (\AA) & $-$47.0 & $-$11 to $-$234 & 7.0\\
  Mass ($M_{\odot}$)& 0.4 & 0.2 to 2.2 & 0.07\\
  Age (Myr)& 0.85 & $\lesssim$\,0.1 to 9 & 0.6\\
  log\,$\dot{M}_{\rmn{acc}}$ ($M_{\odot}yr^{-1}$) & $-$8.1 & $-$6.3 to $-$9.4 & 0.6 \\
\hline\end{tabular}}
\end{table}

\subsection{Caveats}
   
We discuss the differences in our results arising from adopting different stellar evolutionary models, and the effect of setting the binary fraction to zero.
    
\subsubsection{Stellar evolutionary models}
             
The mass and radius used to calculate the mass accretion rates, and the age of the CTTS are determined from interpolating their position in the CMD relative to stellar model tracks and isochrones. Apart from differences due to individual uncertainties in a star's photometry or extinction, the model tracks and isochrones also differ between various authors. This is because of differences in the input physics, in particular the treatment of the birthline, accretion history and the deuterium burning phase. Moreover, the conversion of luminosity to magnitude based on bolometric corrections, and the colour-temperature conversions are highly non-linear, especially for late-type stars, leading to slight differences. 

\begin{figure}   
\center  
\includegraphics[width=80mm, height=120mm]{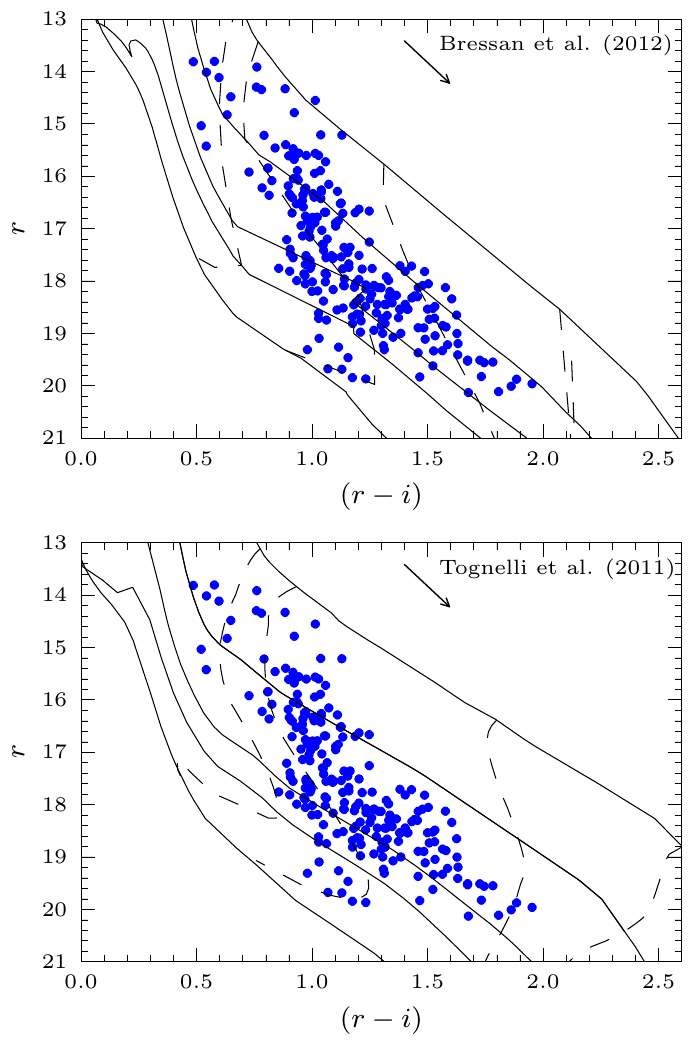}
\caption{The $r$ versus ($r-i$) colour-magnitude of CTTS candidates star overlaid with the Bressan et al. 2012 (top) and Tognelli et al. 2011 (bottom) stellar tracks and isochrones. For both panels, dashed lines from bottom up represent stellar tracks of 0.1, 0.3, 0.6, and 0.8\,$M_{\odot}$, while solid lines from right to left represent PMS isochrones having age 0.01, 0.3, 1, 3, and 10\,Myr respectively.}
\label{fig:complete1}
\end{figure}

\begin{figure} 
\center
\includegraphics[width=80mm, height=135mm]{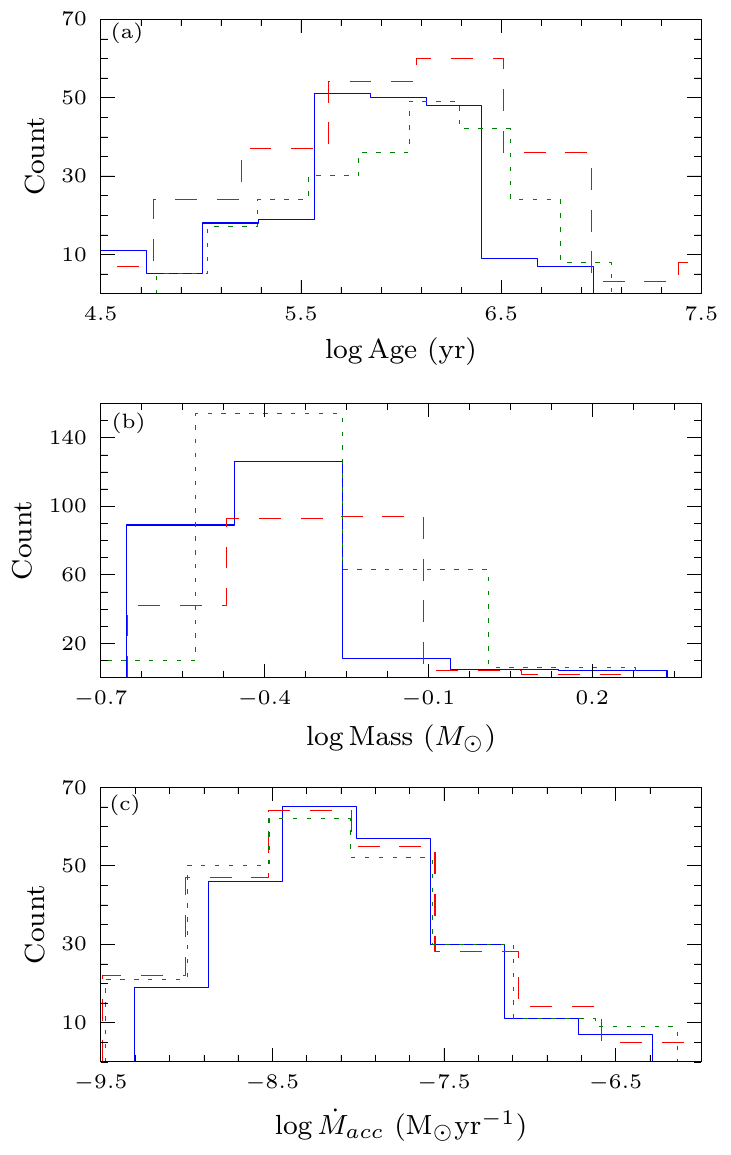}
\caption{The distribution of masses (a), ages (b) and $\dot{M}_{\rmn{acc}}$ (c) determined using different stellar evolutionary models. Solid blue lines are results estimated using the Siess et al. (2000) models, dashed red lines using the Bressan et al. (2012) models and dotted green lines using the Tognelli et al. (2011) models.}
\label{fig:comphist}
\end{figure}

To estimate the uncertainties in the measured properties due to differences between model isochrones and tracks we compare the stellar masses, and ages derived from the Siess et al. (2000) stellar models with the PMS models of \cite*{pisa} and \cite{bress12}. In Fig.~\ref{fig:complete1}, we plot the $r$ vs. ($r-i$) CMD of CTTS overlaid with each of the stellar models. We also plot the histograms (Fig.~\ref{fig:comphist}) showing the estimated distributions of stellar masses and ages interpolated from each of the models compared to the distributions estimated from the Siess et al. (2000) models. 

We find that the median age of our sample determined using Tognelli et al. (2011) isochrones is 1\,Myr and Bressan et al. (2012) isochrones 0.9\,Myr. The distribution of stellar ages derived using the Siess et al. (2000) and Tognelli et al. (2011) isochrones are similar, but there is a larger number of CTTS having ages $>$\,2\,Myr according to the Bressan et al. (2012) isochrones, i.e. evidence for an age spread. This result is similar to that of Sung et al. (2000), where the authors found age spreads ranging between 2 to 5\,Myr using different stellar isochrones. We also point out that the spacing between isochrones decreases from 0.5\,mag in colour for ages 0.1\,-\,2\,Myr to less than 0.5\,mag between 2\,-\,10\,Myr. Small differences in the conversion from the theoretical to the observational plane of the Hertzsprung-Russell diagram may lead to differences in the precision of the interpolated ages. The turnover of the masses determined using the Siess et al. (2000) tracks ($M_{\ast}$\,=\,0.35\,$M_{\odot}$) is slightly lower than those of Tognelli et al. (2011) and Bressan et al. (2012) at $M_{\ast}$\,$\approx$\,0.4\,$M_{\odot}$. In the distribution of Bressan et al. (2012) masses, there is a long tail towards the higher masses. Sung et al. (2000) also report differences in the estimated masses of Lagoon Nebula PMS stars when using different stellar tracks. Finally, we plot the distribution of $\dot{M}_{\rmn{acc}}$ calculated using the stellar masses and radii estimated from the different stellar models (Fig.~\ref{fig:comphist}c). We find that our median log\,$\dot{M}_{\rmn{acc}}$ determined using Bressan et al. (2012) and Tognelli et al. (2011) are smaller by $-$0.05\,dex compared to the Siess et al. (2000) median of log\,$\dot{M}_{\rmn{acc}}$\,=\,$-$8.1\,$M_{\odot}$\,$yr^{-1}$. There is a tail of lower $\dot{M}_{\rmn{acc}}$ sources found using the Tognelli et al. (2011) and Bressan et al. (2012) models, absent in the distribution of Siess et al. (2000) $\dot{M}_{\rmn{acc}}$. The $\dot{M}_{\rmn{acc}}$ estimates are less sensitive to the stellar models.

Finally, we note that a significant fraction of stars with stellar masses $M_{\ast}$\,$>$\,0.6\,$M_{\odot}$ (or $r$\,$\gtrsim$\,16.8) in Fig.\,~\ref{fig:cmd} have comparatively younger ages ($t_{\ast}$\,$<$\,0.8\,Myr) than stars having 0.2\,$M_{\odot}$\,$<$\,$M_{\ast}$\,$<$\,0.6\,$M_{\odot}$, the bulk of whose ages lie between 0.8\,-\,2\,Myr according to the Siess et al. (2000) isochrones. The location of these stars is similar when compared to the \cite{pisa} and \cite{bress12} PMS models (Fig.~\ref{fig:comphist}). It is possible that a bulk of these stars could be foreground reddened late-type stars. However, all stars having $u$-band photometry (53\,per\,cent of the total sample) have demonstrable $u$-band excesses, and 86\,per\,cent lie near the CTTS locus in the near-infrared colour-colour diagram. Evidence of multiple accretion signatures suggests that the number of contaminants in this sample is low. Another observational effect is that in the adopted EW$_{\rmn{H}\alpha}$-spectral type selection criterion, stars spanning F5-M2 spectral types are selected as CTTS if they have EW$_{\rmn{H}\alpha}$ between 10-12 \AA. More massive stars have greater luminosities at similar EW$_{\rmn{H}\alpha}$ leading to systematically higher H$\alpha$ luminosities and thereby accretion rates at higher masses. But, in Section 5 our detection limits calculated based on the EW$_{\rmn{H}\alpha}$-spectral type selection criterion and stellar parameters from a 1\,Myr isochrone indicate that the lack of low accretion rates at $M_{\ast}$\,$>$\,0.6\,$M_{\odot}$ is not likely to be due to the EW$_{\rmn{H}\alpha}$-spectral type criterion. When compared to other CMD of the Lagoon Nebula across a similar mass range, we note a very similar split sequence in age at 0.6\,$M_{\odot}$ in the theoretical colour-magnitude  diagram of Sung et al. (2000; Fig. 9a in that Paper), and a not as prominent effect at $\sim$\,0.8\,$M_{\odot}$ in the CMD of Prisinzano et al. (2005; Fig. 7 in that paper). Isochrones fitted to open cluster may yield good fits to one section of the CMD, but may systematically deviate from the observed sequence in another section in some cases (\citealt*{bon04}; \citealt{mayne07}). Lastly, it is possible that the comparatively more massive stars may have been formed later than the less massive stars, and the observed difference in ages is real. Further information on membership is required to differentiate the two effects.         
     
\subsubsection{Bias due to unresolved binaries}
Due to the unknown dependence of accretion properties on binarity, it is not possible to design a correction for the brightening effect of unresolved/unidentified binaries. We do not apply any corrections due to this effect. 

The angular resolution of our observations (0.5$\arcsec$) implies a resolution of 625\,AU at the distance to the Lagoon Nebula. Our observations are unable to rule out the presence of binaries having separations less than this distance. The effect of undetected binaries in our sample may be understood if we consider the change in the properties of the star if it is physically an equal mass binary system. There is no colour change, while all its magnitudes are 0.75\,mag brighter. This effect is similar to shifting the model tracks to account for differences between different stellar evolutionary models.   

However, estimating the number of possible binary contaminants in our sample is difficult as previous studies of T\,Tauri binaries have been inconclusive. \cite{bouv06} suggest that close binaries may lose their discs significantly faster based on mid-infrared observations, meaning that most disc bearing CTTS are not generally found in close binaries. \cite{daem13} resolved the components of 19 T\,Tauri binaries in Chameleon I. The authors found that $\dot{M}_{\rmn{acc}}$ estimates only for the closest binaries ($<$ 25 AU) were significantly affected.

\subsection{Completeness of our CTTS sample}

The completeness limit of the photometry is estimated from the peak magnitude of the luminosity function. We find that peak magnitudes are approximately 19.8, 18.8 and 19.4\,mag in $r$, $i$, and H$\alpha$ respectively. Using the reddening and distance corrections of Section 4.1, this translates to a limit of about 0.3\,$M_{\odot}$ in the $r$ versus ($r-i$) CMD. This limit can be seen in the histogram of derived masses (Fig.~\ref{fig:hist}b).

In reality, the completeness of the final sample of CTTS is also affected by the EW$_{\rmn{H}\alpha}$ selection criterion. The EW$_{\rmn{H}\alpha}$ selection criterion is designed to exclude interloping chromospherically active late-type stars. As a result, the EW$_{\rmn{H}\alpha}$ selection threshold is larger than the lower limit of EW$_{\rmn{H}\alpha}$ in CTTS (see Sec. 3 and White \& Basri 2003). To estimate the number of CTTS missed due to our adopted selection threshold, we must first estimate the total number of Lagoon Nebula members in the mass range we observe, and then estimate how many of them are accreting, i.e. CTTS. 

The total number of stars in the mass range we observe (0.2\,-\,2.2\,$M_{\odot}$) can be estimated from the initial mass function, provided the slope, maximum stellar mass and total cluster mass of region is known. We adopt the results of Prisinzano et al. (2005), where the authors find a power law index of 1.22$\pm$0.17, and a total cluster mass of $\sim$930\,$M_{\odot}$. The maximum known stellar mass in the region is 20\,$M_{\odot}$ \citep{weidner06}. Based on these results, we estimate that $\sim$ 560, 260, and 40 Lagoon nebula members are present in the 0.2-0.4\,$M_{\odot}$, 0.4-1\,$M_{\odot}$, 1-2\,$M_{\odot}$ mass ranges respectively. \cite{pris07} find that around 40\,per\,cent of member stars in the Lagoon Nebula are actively accreting at a given time, i.e. are expected to be identifiable as CTTS. Combining these results with the estimated number of members, we can estimate the total number of CTTS in the Lagoon Nebula. Comparing this estimated number with our sample, we find that our sample is complete up to 60, 90 and 50\,per\,cent in the 0.2-0.4\,$M_{\odot}$, 0.4-1\,$M_{\odot}$, 1-2\,$M_{\odot}$ ranges respectively. 

\subsection{Contaminants in our CTTS sample}

The positions of the H$\alpha$ identified sample of CTTS in the near-infrared and ($u-g$) vs. ($g-r$) colour diagrams concur with the expected positions of CTTS. The $\dot{M}_{\rmn{acc}}$ measured from H$\alpha$ agrees well with those measured using $u$-band photometry. No stars in our sample have positions beyond the 10\,Myr isochrones, with an overwhelming majority falling around the 1\,Myr isochrone suggesting that background/foreground contamination in our sample is small. This fact, along with the multiple signatures of disc accretion are highly suggestive that the vast majority of our candidate CTTS are genuinely accreting. Moreover, the EW$_{{\rm{H}}\alpha}$ criterion is designed to exclude the expected interlopers such as late type stars whose EW$_{{\rm{H}}\alpha}$ is due to chromospheric activity. However, we do not explicitly select for cluster membership (using for example proper motions). There are only four stars (out of 91) that are accreting in the ($r-i$)\,vs.\,($r-$H$\alpha$) diagram and not accreting in the ($u-g$)\,vs.\,($g-r$) diagram. Neglecting that such a change is due to accretion variability it is possible that these stars may be foreground unreddened late-type stars. Due to the nebulosity, background stars are not expected to be found (see Section 1). 

If we assume that the fraction of H$\alpha$ emission stars not showing any $u$-band excess are field stars, then our contamination rate is $\sim$\,5\,per\,cent. The fact that these stars may  not be accreting is strengthened by the fact that most of these stars do not show atypical near-infrared excess. Note that there is no preferential spatial location for these stars. On the other hand, if we take the number of stars falling at the MS locus in the near-infrared to be contaminants, then our contamination rate $\sim$15\,per\,cent. However, a few stars that posses discs and are still accreting may not show infrared excesses due to variations in inclination angles (Meyer et al. 1997), or because they might be accreting from a debris disc. Moreover, it is difficult to differentiate CTTS using near-infrared colours alone, as the median ($H-K$) colour disc excess from the UKIDSS survey in the near-infrared is only slightly larger than atypical main sequence colours (Lucas et al. 2008). In reality, our contamination rate is most likely somewhere in between $\sim$\,5\,-\,15\,per\,cent, and the total number of contaminants is expected to be around 10-30 per\,cent.

Our estimations suggest that our sample is clean, and complete to a sufficient degree to estimate the statistical properties of the PMS stars within the Lagoon Nebula.

\section{$\dot{M}_{\rmn{acc}}$ as a function of stellar mass}

\begin{figure}
\center
\includegraphics[width=86mm, height=58mm]{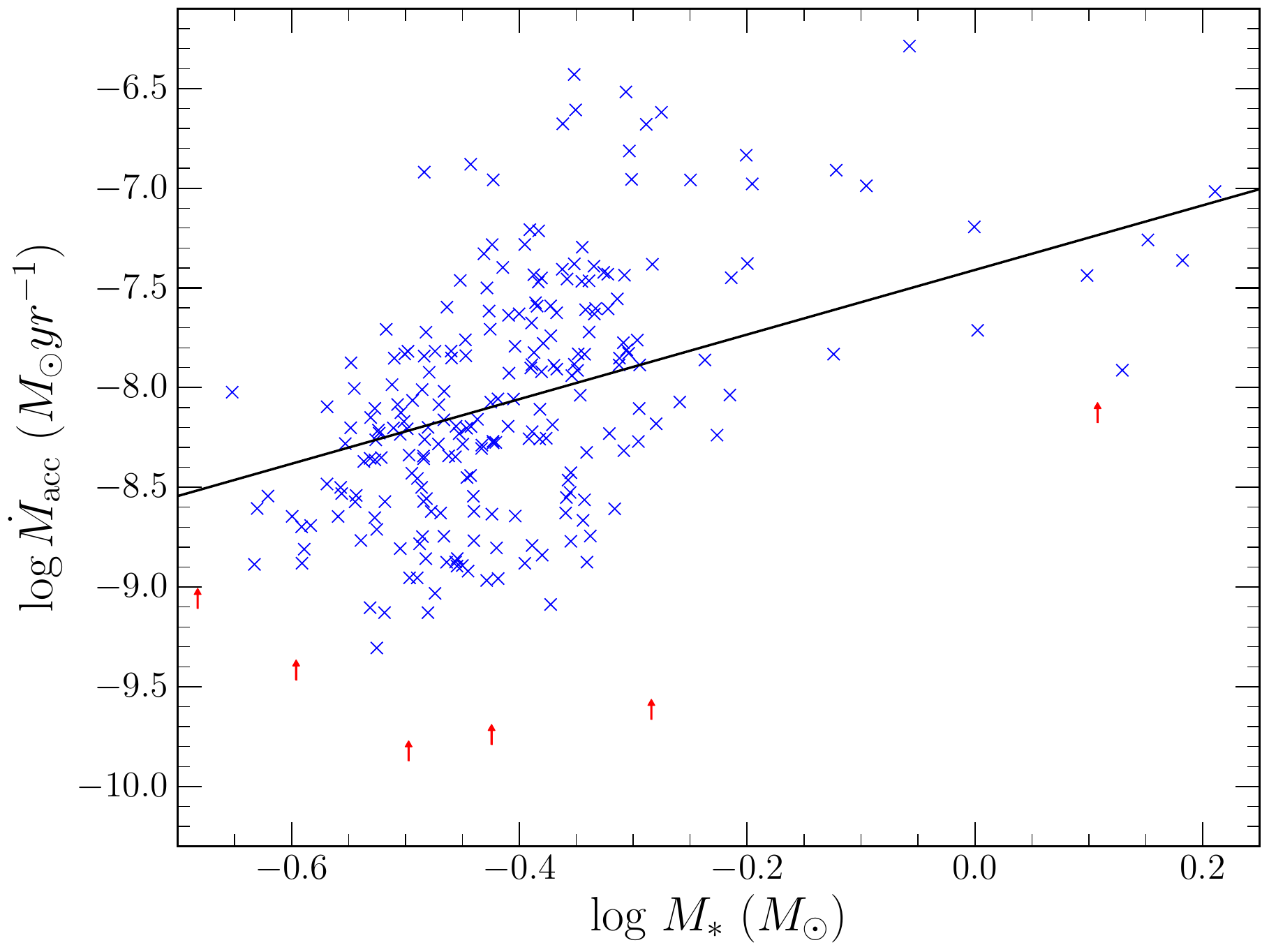}
\caption{The distribution of $\dot{M}_{\rmn{acc}}$ with stellar mass ($M_{*}$) of CTTS is shown by blue crosses. The best-fit power-law is given by $\dot{M}_{\rmn{acc}}$\,$\propto$\,${M_{\ast}}^{2.14\pm0.3}$. Red arrows give the lower detection limits.}
\label{fig:maccm} 
\end{figure}      
 
The $\dot{M}_{\rmn{acc}}$ rate is plotted as a function of ${M}_{*}$ in Fig.~\ref{fig:maccm}. 
We employed survival analysis linear regression using {\sc{ASURV}} \citep{asurv} to account for the lower detection limits in fitting a power law function over the entire mass range. The lower limit of non-detections is calculated adopting the EW$_{\rmn{H}\alpha}$ selection criterion for all points on an 1\,Myr isochrone. We found the best-fit power law index $\alpha$~=~2.14\,\,$\pm$\,\,0.3. Not including the lower limits lead to slightly steeper slope. This strong dependence is well documented in a few star forming regions over a wide range of masses (0.1-3 $M_{\odot}$; see \citealt{muz03, natta06}). The median $\alpha$~$\sim$~1.8-2.2 \citep{hart06}, although much steeper \citep{fang09} or shallower values \citep{geert11} have also been found.       
 
It is important to consider the observational selection effects that may affect the observed relation \citep{clark06, dario2014}. The lack of T\,Tauri candidates at the lower mass end in Fig.~\ref{fig:maccm} is due to the data reaching the detection limit, but we should be able to detect more CTTS at $M_{*}$~$>$~0.5\,$M_{\odot}$ with lower $\dot{M}_{\rmn{acc}}$. Stars with $M_{\ast}$\,$\sim$\,0.5\,$M_{\odot}$ are typically a magnitude brighter than the faint limit. Their paucity cannot be explained by the observational selection limits. Similarly, it has been suggested that the upper envelope of accretion rates at any given mass is claimed to be due to the difficulty of detecting stars with $L_{\rmn{acc}}$\,$>$\,$L_{\ast}$ \citep{clark06}. However, such {\it continuum} stars constitute only a small fraction of the known population of PMS stars (Hartmann 2008). A small fraction of high $\dot{M}_{\rmn{acc}}$ across the mass range will bias the observed relations, but cannot explain it. Therefore, observational limits can bias the observed relation, but cannot easily explain it. 

Overall, there exists a spread greater than an order of magnitude in accretion rates at any given mass in Fig.~\ref{fig:maccm}. In Fig.~\ref{fig:ugr1}, the accretion rates of 87 CTTS from the {\it u} and H$\alpha$ excess emission have a mean difference $\lesssim$ 0.17\,dex, much smaller than the spread in accretion rates. There are more likely other intrinsic factors besides stellar mass controlling accretion rates. Due to the ages of individual stars being uncertain, we do not examine the variation of $\dot{M}_{\rmn{acc}}$ as a function of individual stellar ages.



\section{Spatial distribution of CTTS candidates}

\begin{figure*} 
\center
\includegraphics[width=160mm, height=120mm]{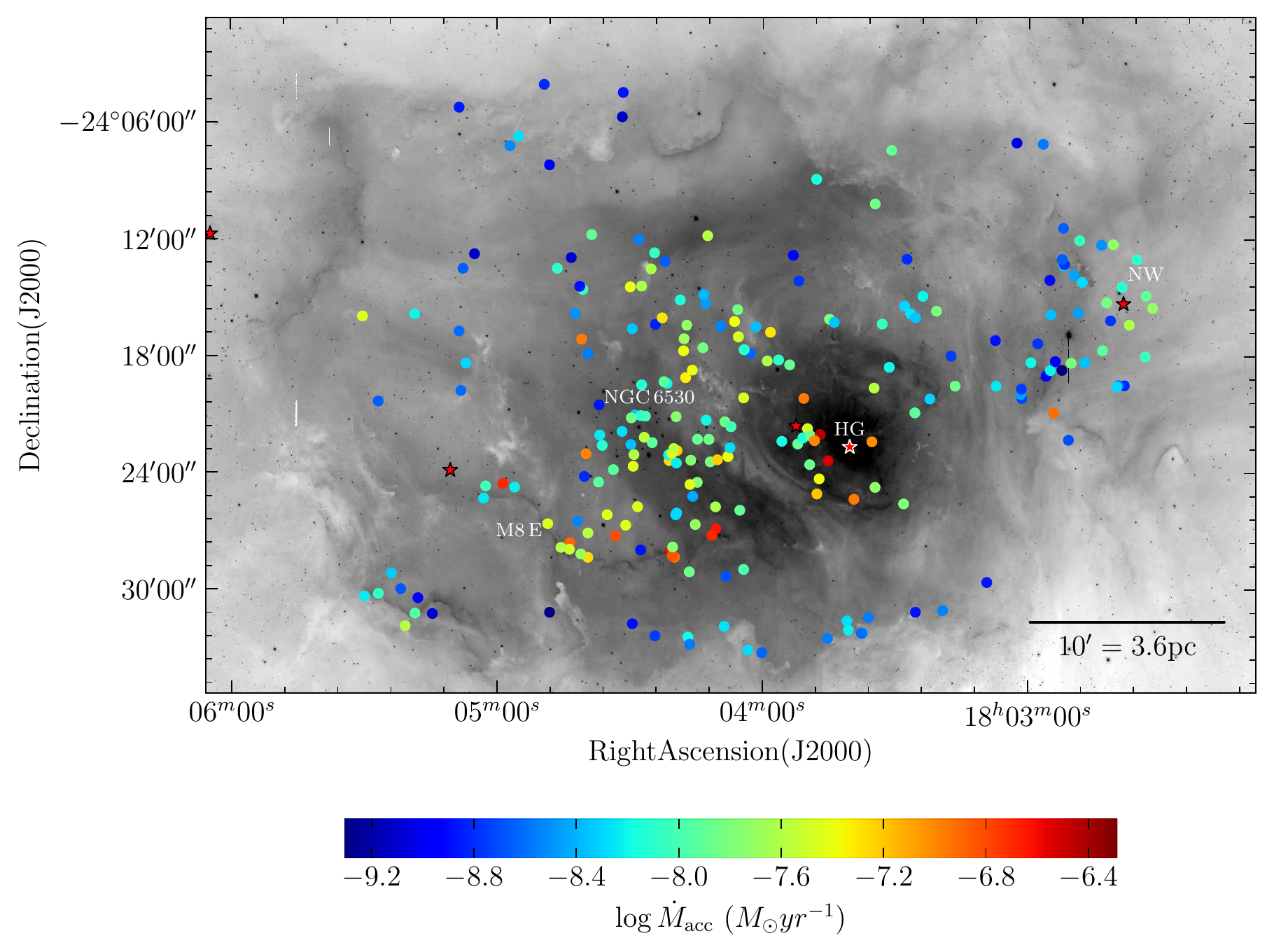}
\caption{CTTS candidates overplotted on the H$\alpha$, shown as inverted grey scale. The plotted objects are coloured according to their estimated mass accretion rates.  The positions of known O-type members are picked out by asteriks. The approximate centres of the Hourglass nebula (HG), NGC\,6530 cluster, the positions of the M8\,E rim and the north-western region (NW) are also labelled. The black bar marks an extent of 3.6\,pc at the adopted distance to the Lagoon Nebula.}
\label{fig:spamacc}
\end{figure*}  

Fig.~\ref{fig:spamacc} shows the spatial distribution of CTTS candidates. Known O-type stars in the field are also shown. CTTS candidates are concentrated in multiple sub-clusters (having density $>$\,10\,stars\,pc$^{-2}$) in four regions:
\begin{enumerate}
\item in the southern bright rim M8\,E. Tothill et al. (2002) identified several gas clumps within this region. A Herbig-Haro object has also been reportedly found by Arias et al. (2006). The O6\,Ve star HD\,165052, which is a confirmed member of the Lagoon nebula \citep{tothill02} lies at the eastern edge of the rim. 
\item the embedded central cluster NGC\,6530, which has a significant number of CTTS located to its immediate north. 
\item the Hourglass nebula region (marked as HG in ~\ref{fig:spamacc}), and a region to its north east. It contains at least two O-type members, the O4\,V star 9\,Sgr, and the O7\,V star H\,36 which are thought to be the chiefly responsible for ionising the Lagoon and Hourglass nebula respectively. 9\,Sgr has 10 CTTS identified within a 1\,pc radius. The number of CTTS surrounding H\,36 is thought to be similar (see Arias et al. 2006), but cannot be confirmed in our study due to the difficulty in subtracting the highly variable background around H\,36. Both \cite{white97} and Tothill et al. (2002) report a significant amount of molecular gas (10-30\,$M_{\odot}$) still shielding H\,36, forming a compact H\,II region inside the molecular cloud. 
\item a north-western region (marked as NW in ~\ref{fig:spamacc}) at the edge of the visible nebulosity, surrounding the O9 star HD\,164536, which is not a confirmed member. This region has received comparably little interest in previous studies of the Lagoon Nebula, owing to the brighter core, where star formation appears to be most active.  
\end{enumerate}

A small fraction ($\lesssim$\,6\,\%) of relatively more isolated stars are situated along the edges of the Lagoon Nebula, wherein some cases the nearest CTTS candidate is located at distances $\gtrsim$\,1\,pc. The overall morphology of our sample is indicative of a continuum ranging from dense clustering to comparative isolation. This is similar to results in nearby star-forming regions \citep{allen07}. Our spatial over densities of CTTS candidates are very similar to those found by \cite{kuhnmystix}.

\subsection{Distribution of accretion rates}

It is interesting to see whether the spatial morphology of CTTS candidates is reflected in the distribution of accretion rates, and if any inference of how star formation has proceeded can be drawn from it. \cite{lada76} proposed star formation proceeded east to west through time beginning in NGC\,6530 which triggered star formation in the Hourglass nebula. \cite{lightfoot84} put forth a similar scenario, where star formation was triggered by the NGC\,6530 cluster in the Hourglass nebula region, and cite the presence of peripheral O stars in the Lagoon nebula as evidence for a younger population. Alternatively, Damiani et al. (2004); Prisinzano et al. (2005) and Prisinzano et al. (2012) suggested that star formation progressed from the north to the south, and the north-eastern CTTS were formed initially and have currently shredded their molecular gas, and star formation proceeded towards the west and south-west. Arias et al. (2007) reported that the ages of CTTS decreased from the north to the south. Tothill et al. (2008) consider these results, and suggest that star formation may have proceeded outwards from the core, evidenced by age gradients found by Damiani et al. (2004) and Arias et al. (2007), and the young sources around the dense molecular cloud cores at the edges of the nebula.    

\begin{figure*}
\center
\includegraphics[width=100mm, height=182mm]{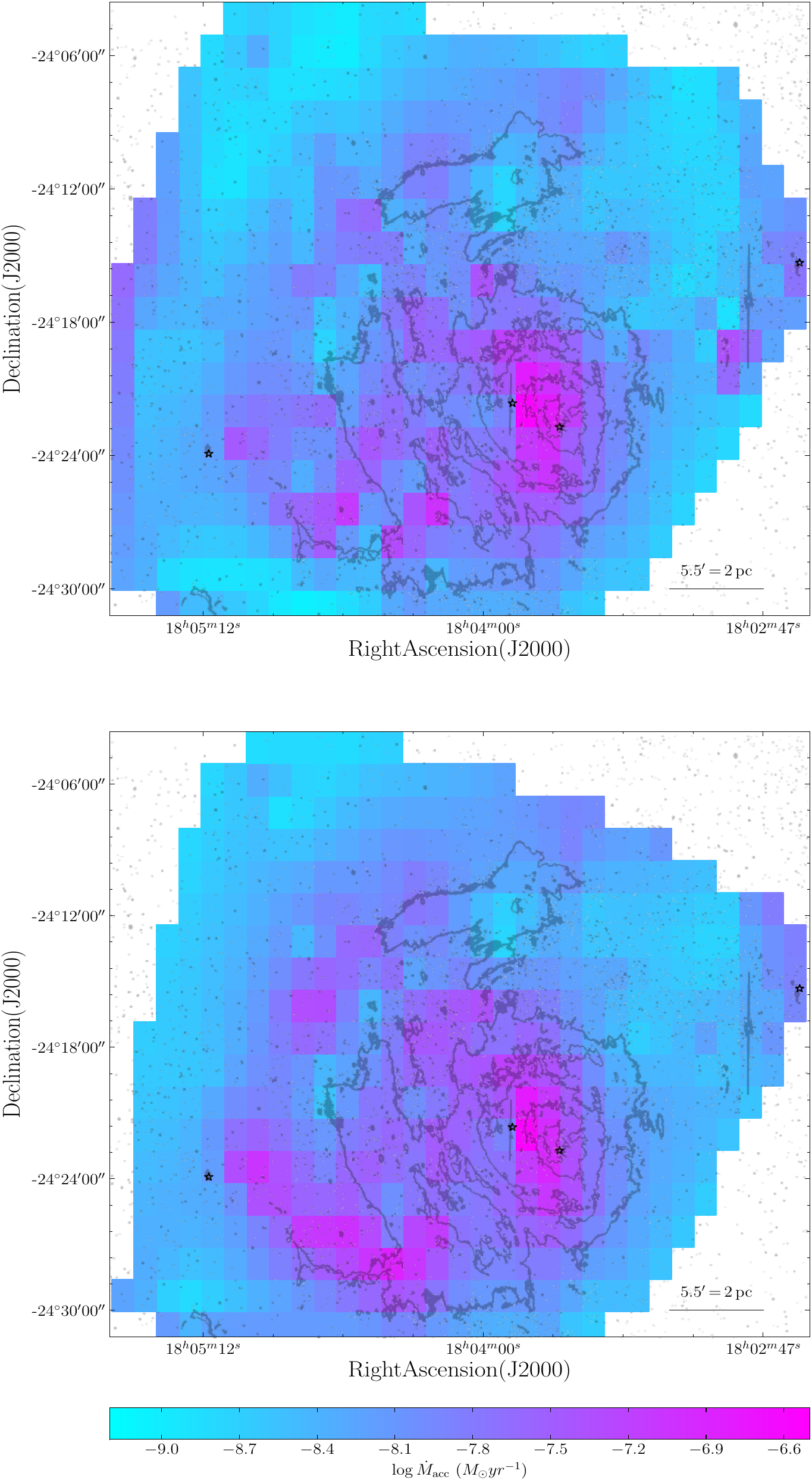}
\caption{Density distribution of $\dot{M}_{\rmn{acc}}$ in the complete mass range (top) and in the restricted range 0.35\,$M_{\odot}$\,$<$\,$M_{\ast}$\,$<$\,0.6\,$M_{\odot}$ (bottom), which contains 135 stars. The colour is representative of the median $\dot{M}_{\rmn{acc}}$ within a 0.5\,pc$^{2}$ box. White is regions with no identified CTTS. The corresponding colourbar is shown at the bottom. Known O-type stars (asteriks) are shown for reference. The approximate centres of the Hourglass nebula, NGC\,6530 cluster, and M8\,E region are annotated following Fig.~\ref{fig:spamacc}. Overplotted are H$\alpha$ image contours at 1\%, 5\%, 10\% of the total emission, which outline the diffuse nebular emission.}
\label{fig:maccagespat}
\end{figure*}

To investigate such scenarios, a 2D histogram of accretion rates covering the entire mass range, and restricted to the 0.35\,$M_{\odot}$\,$<$\,$M_{\ast}$\,$<$\,0.6\,$M_{\odot}$ are shown in Fig.~\ref{fig:maccagespat}. The colour is representative of the median accretion rate of all the stars situated in a 0.5\,pc$^{2}$ box. The diffuse H$\alpha$ contours, which outline the nebular emission are also shown. The distribution of the accretion rates is considered to be due to the star formation history of the Lagoon nebula. Younger stars with higher accretion rates are thought to be located closer to their birth locations, while older accreting CTTS may have had time to dynamically evolve and move away from their natal molecular cloud which is thought to be reflected by larger spacing between them, and lower accretion rates. However, the accretion rate also depends steeply on stellar mass (see Fig.~\ref{fig:maccm}), which is why we corroborate our map covering the entire observed stellar mass range with one restricted to 0.35\,-\,0.6\,$M_{\odot}$ CTTS, which includes 135 candidates. Based on the best-fit from Fig.~\ref{fig:maccm}, stellar mass affects the $\dot{M}_{\rmn{acc}}$ by $\sim$\,0.4\,dex in this range, so any larger variations of $\dot{M}_{\rmn{acc}}$ are likely reflective of dependences on stellar ages. Therefore, the distribution of accretion rates can be an indicator of the star formation history within the region, and also reflective of any intrinsic age spread. 

In Fig.~\ref{fig:maccagespat}, we find that the median accretion rates are highest in the M8\,E and Hourglass region ($-$7.2\,$M_{\odot}$yr$^{-1}$), and decrease non-uniformly towards the edges of the nebula. The accretion rate of the central embedded cluster NGC\,6530 is between $-$7.5 to $-$8.2\,$M_{\odot}$yr$^{-1}$, and the north-western region fractionally lower ranging between $-$7.8 to $-$8.2\,$M_{\odot}$yr$^{-1}$.

Our results do not corroborate completely with any single sequential star formation scenario proposed in the literature. If, we consider star formation to have proceeded east to west as described by Lada et al. (1976), why are the median accretion rates in the south-eastern rim so high? Also, the region is known to contain at least one Herbig-Haro object (see Tothill et al. 2008), and dense CO clumps suggesting youth, if not younger then at least similar to the age of the Hourglass nebula region. Alternatively, if accretion proceeded north to south as suggested by Prisinzano et al. (2005; 2012), the median accretion rate of the north-western region is comparable to the central NGC\,6530 cluster located to its south. The north-western region has not been as comprehensively studied as the central Nebula, but \cite{dewa10} identify six Class\,I objects using {\it{Spitzer}} infrared observations near the O-type star HD\,164536, and comparable to the number density of Class\,I objects found in the NGC\,6530 cluster. This gives weight to Fig.~\ref{fig:maccagespat} that accretion in the north-eastern edge of the Lagoon Nebula is proceeding at rates fractionally lower to the central cluster. Alternatively, if star formation proceeded outwards from the core of NGC\,6530 as suggested by Tothill et al. (2008) why are older stars found towards the northern, and eastern edges, but not in the southern and western edges? It is possible that low $\dot{M}_{\rmn{acc}}$ CTTS in regions of comparatively higher values of extinction are not detected in the study, which could lead to observed spatial dependencies. But the overall spread in measured extinction values within the region is small ($A_{V}$\,$\sim$\,0.5\,mag, see Section 1). In our mass restricted $\dot{M}_{\rmn{acc}}$ map, we find similar dependencies even though the imposed limit of 0.35\,$M_{\odot}$ places any star with extinction $A_{V}$\,$<$\,2\,mag within our detection limits. Both these considerations suggest that differential extinction is unlikely to bias our results significantly. Our results lead us to suggest that a simple sequential star formation process may be too simplistic to apply throughout the Lagoon Nebula, although there is evidence that some regions of the Lagoon nebula may be older than others. Such a conclusion was discussed briefly in \cite{getman} in their multi-region study conducted using archive infrared and X-ray data. Instead, the distribution of accretion rates may be reflective of the clumped nature of the collapsing natal molecular cloud and the proximity of ionising stars. The reasons for cloud collapse (or {\it{triggering}}) of star formation may be to due to multiple factors, and therefore not explained by a single scenario.

\subsection{Mapping of CTTS and protostellar populations}

\begin{figure*}
\center
\includegraphics[width=160mm, height=120mm]{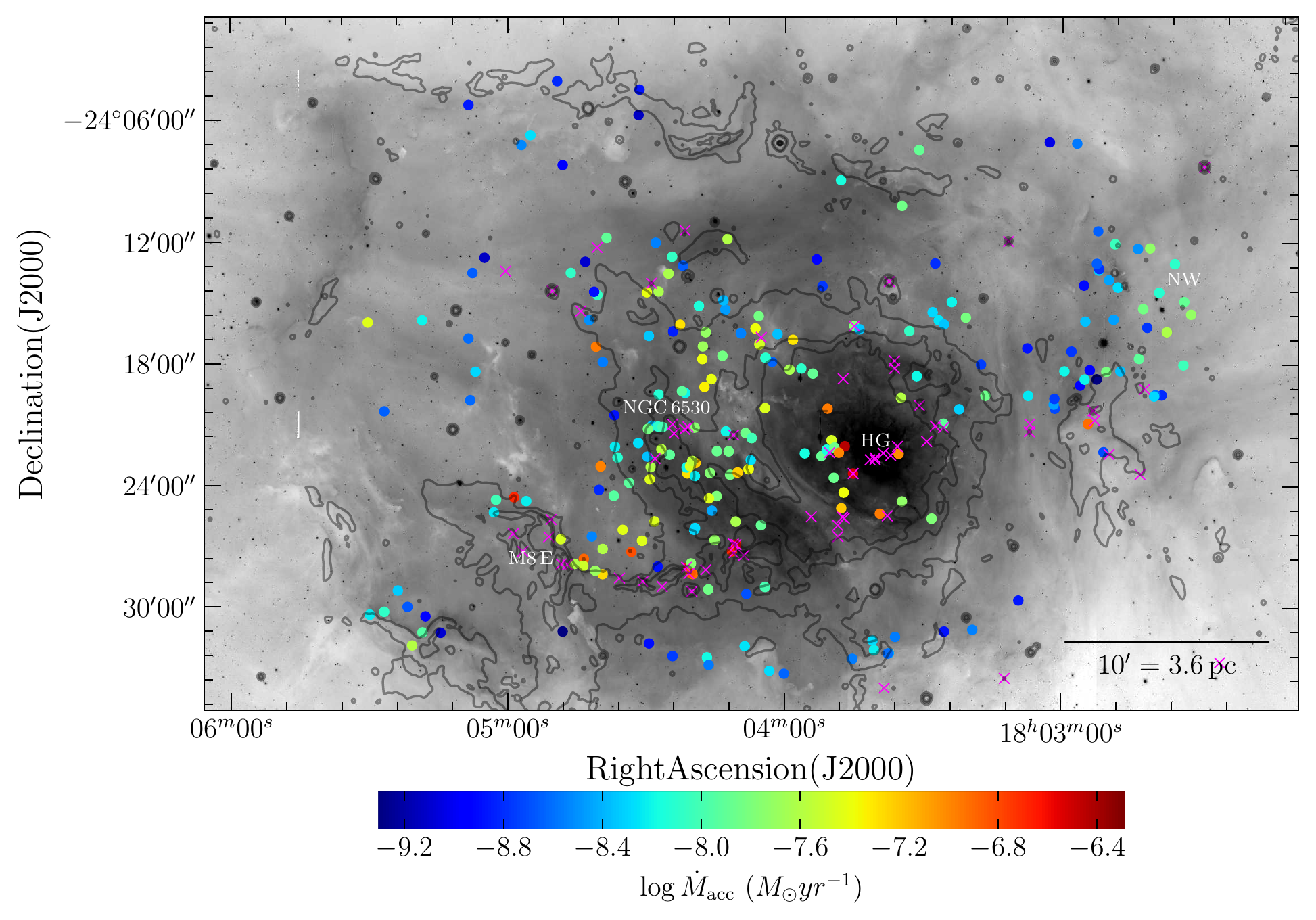}
\caption{Distribution of CTTS candidates (circles), and Class\,I objects (magenta crosses) identified by Kumar \& Anandrao (2010) overlaid on {\it{Spitzer}} 8$\micron$ contours. Colour of the circles is representative of measured log\,$\dot{M}_{\rmn{acc}}$, as in Fig.~\ref{fig:spamacc}. The approximate centres of the four regions are annotated following Fig.~\ref{fig:spamacc}.}
\label{fig:proto}
\end{figure*}

\begin{figure*} 
\center
\includegraphics[width=120mm, height=50mm]{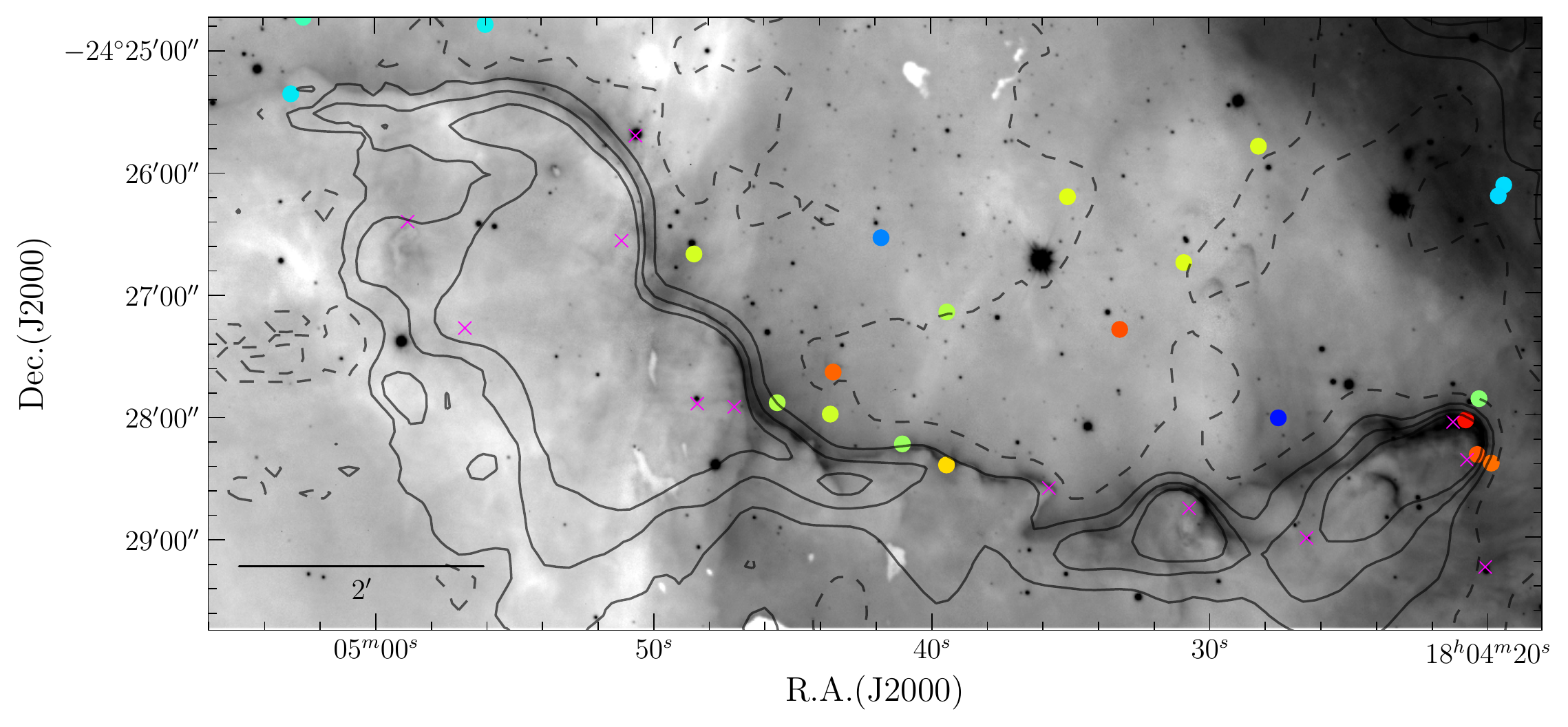}
\caption{Inverted H$\alpha$ greyscale of the M8\,E region overlaid by {\sc{SCUBA}} 850$\micron$ contours. Symbols are same as Fig.~\ref{fig:proto}. The O-type stars 9\,Sgr and H\,36 are located to the northwest.}
\label{fig:image}
\end{figure*}  

The observed spatial distributions of young stellar objects (YSO) in star forming regions are a relic of the star formation process, and also indicative of their dynamical evolution from the protostellar to the pre-main sequence phase. Recent surveys of nearby star-forming regions (for example, see \citealt{win07}; \citealt*{meg09}) have found young Class I protostars densely packed and tracing their natal molecular clouds, while relatively older pre-main sequence stars are found to be comparatively more spread out. Such morphologies place useful constraints on current star formation theories (\citealt{meg09, moe09, hig13}), and can also be useful diagnostics of the evolutionary stage of PMS stars. 

The sample of Class\,I objects is drawn from a {\it{Spitzer}} study of the region by \cite{dewa10}. Briefly, the authors of that paper used observations in the {\it{Spitzer}} 3.6, 4.5, 5.8 and 8\,$\micron$ bands to classify PMS objects according to their spectral slope. They identified 65 Class\,I objects on this basis. The distribution of our VPHAS+ CTTS sample, and Class\,I sources from \cite{dewa10} are over-plotted on contours of {\it{Spitzer}} 8\,$\micron$ emission in Fig.~\ref{fig:proto}. The region surrounding the Hourglass nebula is saturated in the {\it{Spitzer}} image. Contours display a central egg shaped morphology, with elongated filamentary structures found across the northern, southern and western edges. Class\,I objects are found within the filamentary gas structures indicated by the contours. CTTS are found along the edges of the contour lines, and appear to {\it{hug}} the gas structures, with extended distributions in halos surrounding the Class\,I objects.   

The distribution of the nearest neighbours of Class\,I objects has a median projected distance of 0.18\,pc. The CTTS median nearest neighbour separation is 0.24\,pc. There are no defined peaks, but an exponential decrease after a gradual peak. On performing a Kolmogorov-Smirnov (K-S) test, we find the probability that Class I and Class II objects are derived from the same random distribution to be ~\,10\,\%, i.e. non-negligible (p-value\,=\,0.09). We also generate a random distribution of 10,000 stars, where no underlying distribution is assumed and we calculate their nearest neighbour separations. Comparing the YSO distribution with the sample of random stars, the p-value in both cases is $<$0.01\,\%. The K-S tests suggest that the Class I objects and CTTS are similar. Both cannot be drawn from a randomly distributed population of stars, suggesting significant clustering. We caution that our numbers may be biased as both the CTTS and Class\,I samples are selected from data which are affected in the Hourglass nebula, thus possibly not identifying important sources within this region. 

\begin{figure}  
\centering
\includegraphics[width=80mm, height=59mm]{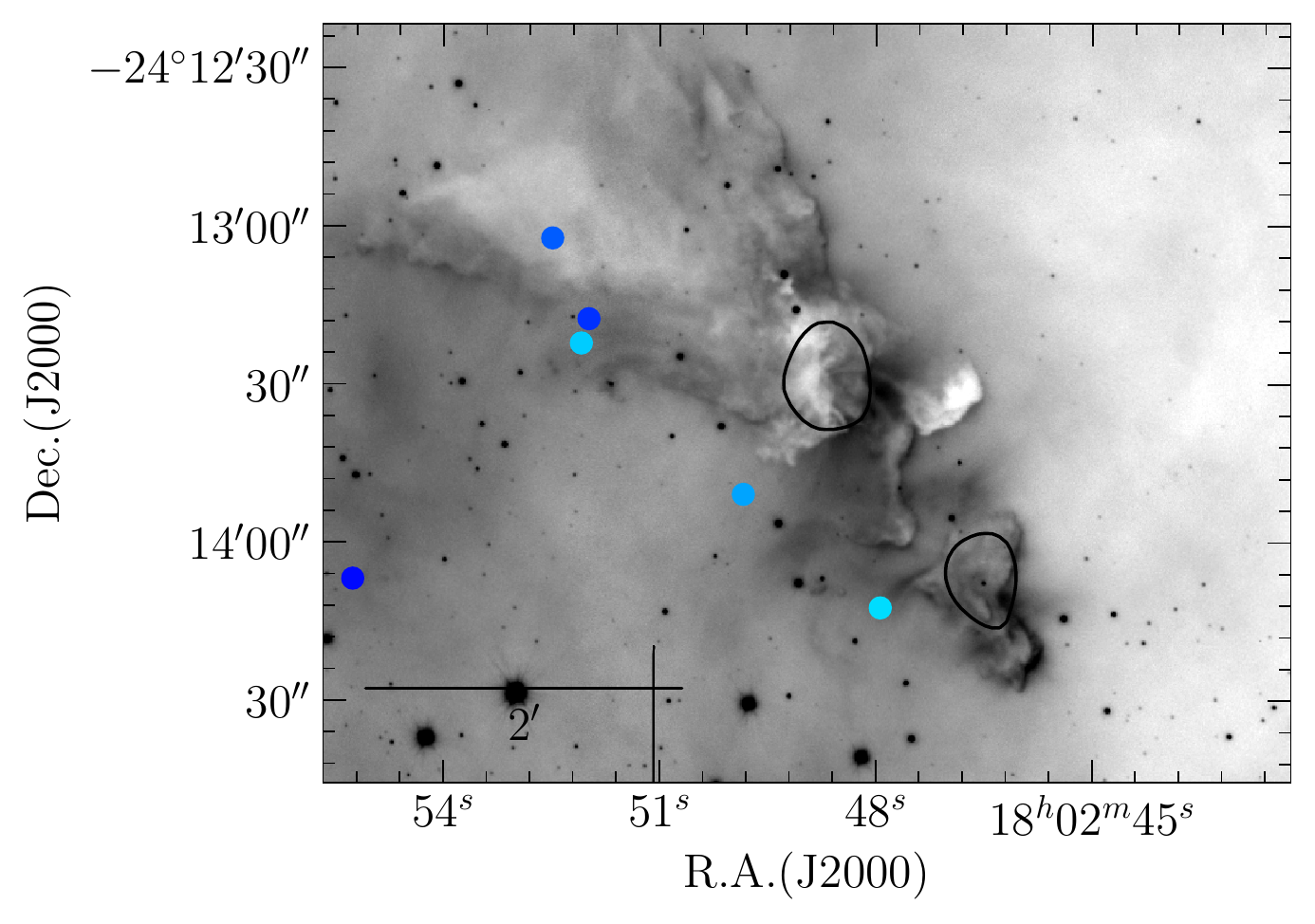}
\caption{Inverted H$\alpha$ greyscale image of an elephant's trunk like structure in the Northwest of the Lagoon nebula region overlaid by {\it{Spitzer}} 8$\micron$ contours. Symbols are same as Fig.~\ref{fig:proto}. The O-type star HD\,164536 is located to the south west, and is possibly ionising the western rims of the structure.}
\label{fig:imageb}
\end{figure}

Adopting the mean molecular cloud properties of the south eastern gas clumps derived by Tothill et al. (2002), where the mean molecular density is 4.46\,$\times$\,10$^{4}$\,cm$^{-3}$, and mean gas temperature around 25\,K we find the Jeans length \citep{jeans28} to be 0.15\,pc. The median observed spacing of Class\,I objects is roughly similar to the Jeans length. Similar results were found in nearby star-forming regions by (\citealt{young06, win07, enoch09}). However, a number of Class\,I objects are found within the median spacing. This has lead \cite{win07} to suggest the possibility of Bondi-Hoyle or competitive accretion in protostars in such regions due to the possibility of overlapping envelopes.

Another important morphology to consider here is the location of the YSO with respect to the ionising stars. In Fig.~\ref{fig:image}, 850\,$\micron$ {\sc{SCUBA}} contours spanning the southern filamentary arc (Tothill et al. 2002) are overlaid on the H$\alpha$ image. Protostars are located within the molecular gas clumps, while CTTS are situated at the edges. The region is being ionised by the O4\,V star 9\,Sgr, and possibly the H\,36 07\,V star located inside the compact H{\scriptsize{II}} region to the north west of the region. The position of the Class\,I objects and CTTS is suggestive of a decreasing age gradient north to south with respect to the central cluster NGC\,6530 and the Hourglass nebula.  

We identify an elephant's trunk like structure in the north west (Fig.~\ref{fig:imageb}). CTTS are located at the outer edges of the structure. The structure may be being ionised by the O\,type star HD\,164536 situated around 1\,pc to its south west, causing the bright rims on the western side of the structure. Most of the molecular gas in the structure has been evaporated, as suggested by the low intensity of {\it{Spitzer}} contours. Compared to the M8\,E feature, the CTTS here are located on the other side of the bright rim being ionised, which is a curious morphology. No Class\,I sources have been identified within the region by Kumar \& Anandarao (2010). CTTS all have an age between 1.6 and 2.3\,Myr. The median radial velocity of members within the region (0.6 pc\,Myr$^{-1}$; Prisinzano et al. 2007), suggests that stars are located close to where they formed. The difference in the morphology of Class\,I YSO and CTTS between these two regions suggests that there is a lack of evidence to apply a global star formation scenario throughout the Lagoon nebula.

\subsection{Accretion properties with respect to O stars}

\begin{figure*} 
\center
\includegraphics[width=170mm, height=78mm]{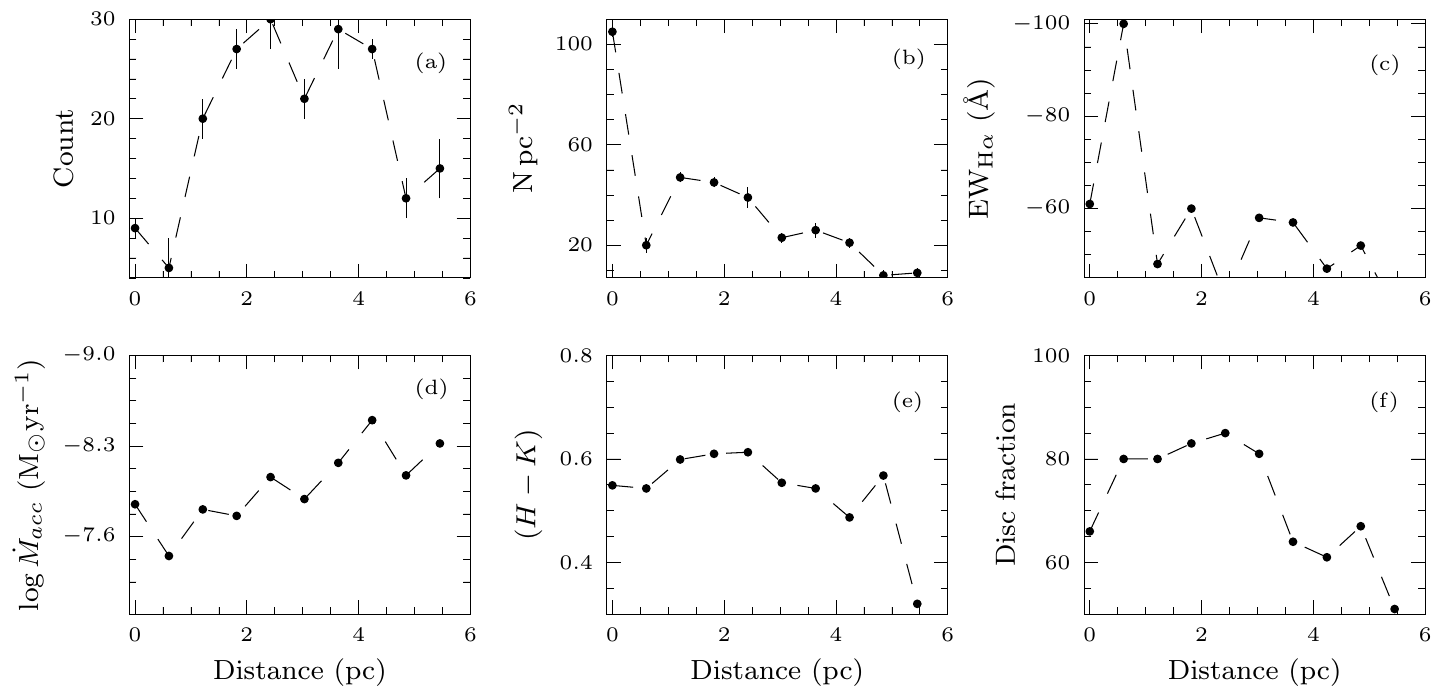}
\caption{Histogram showing the radial variation of number, number density, median EW$_{{\rmn{H}}\alpha}$, median log\,$\dot{M}_{\rmn{acc}}$, median ($H-K$) colour, and disc fraction with respect to the O-type star 9\,Sgr (a-f).}
\label{fig:OB}
\end{figure*}  

\begin{figure*} 
\center
\includegraphics[width=170mm, height=78mm]{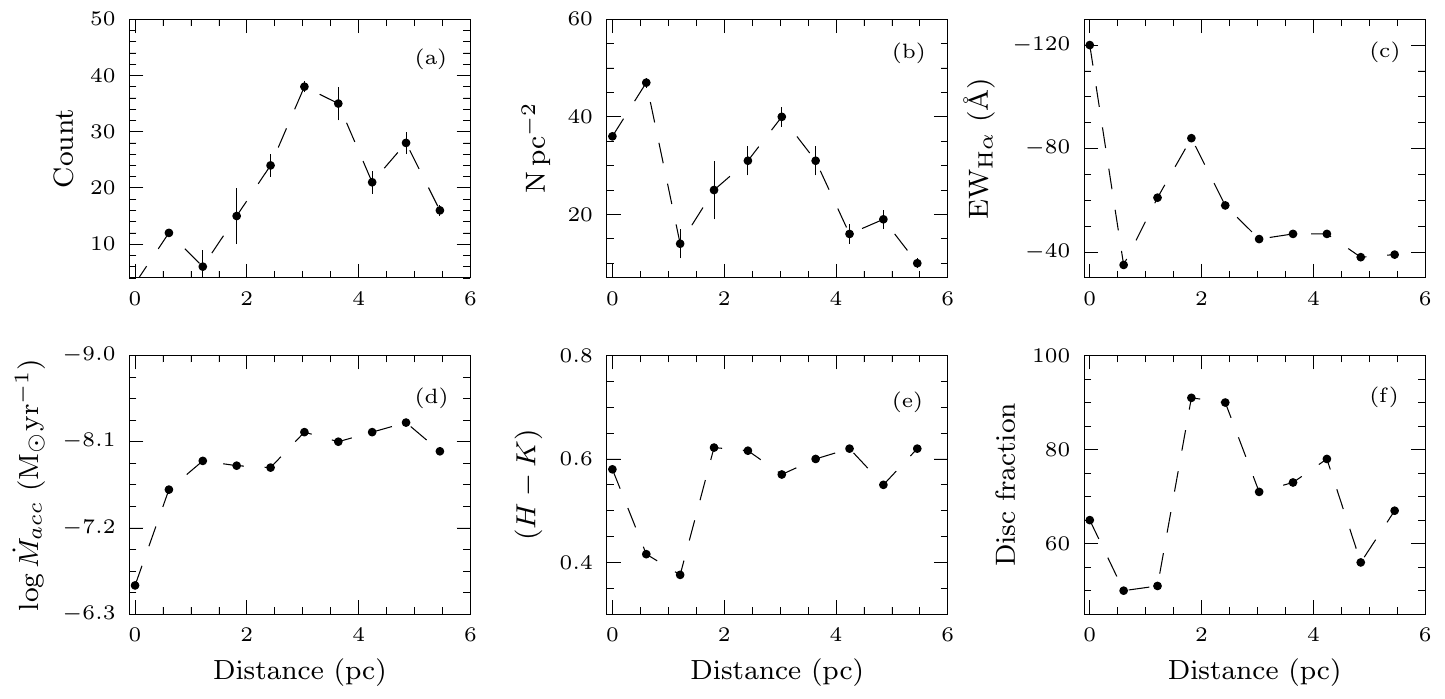}
\caption{Histogram showing the radial variation of number, number density, median EW$_{{\rmn{H}}\alpha}$, median log\,$\dot{M}_{\rmn{acc}}$, median ($H-K$) colour, and disc fraction with respect to the O-type star H\,36 (a-f).}
\label{fig:OB1}
\end{figure*}   

PMS accretion is subject to the influence on its environment by nearby ionising stars. Ionising radiation from O-type stars is thought to disperse nearby molecular gas, and remove circumstellar material from PMS accretion discs disrupting star formation. There has been evidence for the photoevaporation process decreasing the disc fraction of PMS stars within a few parsecs (\citealt{gaur10, roc11}). Alternatively, ionising fronts of O-stars are thought to trigger star formation by collapsing molecular clouds (see \citealt{elm00, sicilia10, geert11}; \citealt*{karr09}; \citealt{snider09}).

To study the effects of O-type stars on the PMS population in the Lagoon nebula, we analyse the radial distributions of the CTTS number density, accretion and disc properties with respect to their distance from the known O-type members in Figs.~\ref{fig:OB} and \ref{fig:OB1}. Two O-type stars are considered, namely the O4\,V type star 9\,Sgr, which is thought to be ionizing the central part of the Lagoon Nebula, and responsible for clearing gas in the rift within the NGC\,6530 cluster (Tothill et al. 2008), and the O7\,V type star H\,36 which is thought to be the main ionizing force in the Hourglass nebula (Arias et al. 2007).

Fig.~\ref{fig:OB} indicates that the number density of CTTS within 1\,pc of 9\,Sgr is considerably higher, with the Spearmans $\rho$\,=\,$-$0.72. The median EW$_{{\rmn{H}}\alpha}$ and log\,$\dot{M}_{\rmn{acc}}$ are also considerably different from the mean within 1\,pc of 9\,Sgr (where $\rho$\,is 0.68 and -0.84 respectively), and follow a steady distribution 2$\sigma$ within the median thereafter. We find little change in the median ($H-K$) value with radial distance from 9\,Sgr except at distances $>$5\,pc, at the edge of the nebula. The CTTS disc fraction (defined as the ratio of CTTS falling within the near-infrared excess region of \citep{lada00} to the whole CTTS population within the region) follows a similar trend, with the disc fraction somewhat smaller within 1\,pc but steady thereafter until beyond 5\,pc.

We find that the number density is fractionally higher within 1\,pc of H\,36 (Fig.~\ref{fig:OB1}), although this region is affected by incorrect H$\alpha$ background sky subtractions (see Section 2). If the number of known CTTS (from Arias et al. 2007) within the excluded region is considered, the number density increases to 72\,\%. We find the median median EW$_{\rmn{H}\alpha}$ and log\,$\dot{M}_{\rmn{acc}}$ are higher within 1\,pc. The disc fraction within 1\,pc is significantly smaller, although we caution that our observations are biased in this region (see Section 2.1). Arias et al. (2006) used near-infrared observations to find that the fraction of disc-bearing members to non-members is around 70\,\% within this region, suggesting that it is close to the median in the Lagoon Nebula. 

Overall, we find that the gradients of CTTS accretion properties with respect to the O-type stars are not indicative of triggered star formation. We find  evidence for disc photoevaporation within 1\,pc of 9\,Sgr and H\,36, although we caution that taking into account censored observations, this may not be the case around H\,36. Finally, we find that the number density, median EW$_{\rmn{H}\alpha}$ and log\,$\dot{M}_{\rmn{acc}}$ are comparatively higher within 1\,pc of both 9\,Sgr and H\,36. This may be explained by the fact that the youngest stars are situated close to the O-type stars, as shown by Fig~\ref{fig:maccagespat}.

\section{Summary}

We identify 235 Classical T\,Tauri candidate stars in the Lagoon Nebula and measured their accretion rates. The majority of our sample demonstrate multiple indicators of disc accretion: H$\alpha$ emission, $UV$-excess, and near-infrared excesses. The conclusions drawn from our study are:

\begin{enumerate}
\item A majority of the CTTS in the Lagoon Nebula have isochronal ages around 1\,Myr. This is consistent with the main-sequence lifetimes of O-type star members, the dynamical evolutionary stage, and the lifetime of the HII region.
\item The accretion rates determined from H$\alpha$ line luminosity and UV-excess luminosity from 87 stars show
no statistically significant differences. The mean amplitude variation $\sim$\,0.17\,dex.
\item We find best-fit power law relation $\dot{M}_{\rmn{acc}}$\,$\propto$\,${M_{\ast}}^{2.1\pm0.3}$.
\item The distribution of CTTS is a continuum ranging from dense clustering to relative isolation. Class I objects are comparatively more clustered than CTTS.
\item The youngest accretors are concentrated around the Hourglass nebula and M8\,E region. Accretion rates decrease towards the edges of the nebula, except in the northwest.
\item We find no evidence supporting a global star formation scenario within the Lagoon nebula. We suggest that star formation may be influenced by environmental effects on smaller scales. 
 
Overall, we demonstrate the application of VPHAS+ data to identify and measure accretion rates of CTTS candidates using uniform selection criteria and known detection limits. Future astrometric and spectroscopic observations such as those undertaken by the GAIA and GES survey will obtain further information about a significant number of our CTTS sample, complementing our VPHAS+ study.   

\end{enumerate}

\section*{Acknowledgements}
Based on observations collected at the European Southern Observatory Very Large Telescope in programme 177.D-3023(C). V.M.K. and J.S.V. acknowledge funding from DCAL-NI and the UK Science and Technology Facilities Council. V.M.K. thanks the anonymous referee, and Drs. R. Oudmaijer and G. Ramsay for providing helpful and constructive comments.
\bibliography{1}{}

\bsp

\label{lastpage}

\end{document}